\documentclass[aps,pra,twocolumn,showpacs,nofootinbib,superscriptaddress,groupedaddress,a4paper]{revtex4-1}
\usepackage{amsmath}
\usepackage{epsfig}
\usepackage{graphicx}
\usepackage{color}
\usepackage{amssymb}
\usepackage{epstopdf}
\usepackage[utf8]{inputenc}
\usepackage[T1]{fontenc}
\usepackage[colorlinks=True, linkcolor=blue, citecolor=blue]{hyperref}
\usepackage{ulem}
\usepackage{lineno}
\begin{document}
\title{Stability of a Bose condensed mixture on a bubble trap}
\author{Alex Andriati$^{1}$}\thanks{andriati@if.usp.br}
\author{Leonardo Brito$^{1}$}\thanks{brito@if.usp.br}
\author{Lauro Tomio$^{2}$}\thanks{lauro.tomio@unesp.br}
\author{Arnaldo Gammal$^1$}\thanks{gammal@if.usp.br}
\affiliation{$^{1}$Instituto de F\'{i}sica, Universidade de S\~{a}o Paulo, 05508-090 S\~{a}o Paulo, Brazil.\\
$^{2}$Instituto de F\'isica Te\'orica, Universidade Estadual Paulista, 01156-970 S\~ao Paulo, SP, Brazil.}
\date{\today}
\begin{abstract}
Stability and dynamical behavior of binary Bose-Einstein condensed mixtures trapped on the surface of 
a rigid spherical shell are investigated in the mean-field level, exploring the miscibility with and without 
vortex charges,  considering repulsive and attractive interactions. In order to compute the critical points 
for the stability, we follow the Bogoliubov-de Gennes method for the analysis of perturbed solutions, with 
the constraint that initially the stationary states are in a complete miscible configuration.
For the perturbed equal density mixture, of a homogeneous uniform gas and when hidden vorticity
is verified, with the species having opposite azimuthal circulation, we consider
small perturbation analysis for each unstable mode, providing
a complete diagram with the intra- and inter-species interaction role on the stability of the
miscible system. Finally, beyond small perturbation analysis, we explore the dynamics of some 
repulsive and attractive inter-species states by full numerical solutions of the time-dependent 
Gross-Pitaevskii equation.
\end{abstract}
\maketitle
\date{\today}
%\pacs{0um3.75.Lm, 67.85.De}
%\keywords{Bose-Einstein condensates, binary condensate, bubble trap, 
%vortices, and topological excitations}
\section{Introduction}\label{sec1}
The reports on the realization of the long-time predicted Bose-Einstein
condensation~\cite{1924Bose,1925Einstein},
with ultra-cold repulsive~\cite{1995Anderson,1995Davis} and attractive interacting atoms~\cite{1995Bradley},
followed by the possibilities to control the atomic interactions via Feshbach
resonance mechanisms~\cite{1958Feshbach} (reported in Refs.~\cite{1998Inouye,Timmermans1999,2010Chin}),
have opened the door to laboratory investigations to probe plenty of quantum phenomena expected to happen
close to zero temperature. Concerning the experimental and theoretical progress on studies with ultra-cold 
gases, some review papers and textbooks~\cite{1999Dalfovo,2003Pitaevskii,2008Pethick,2008Bloch,2009Fetter,2016Pitaevskii} 
are available, providing a broad perspective of the theme, from which other relevant works can be traced.

Just after the first cold-atom experiments, following a theoretical prediction in Ref.~\cite{1996Ho}, two overlapping condensates 
with spin states of $^{87}$Rb was produced in Ref.~\cite{1997Myatt}, with the separation dynamics of the two-spin components 
of the mixture reported in Ref.~\cite{1998Hall}. At this time, the properties of binary condensed mixtures having been also 
investigated in Ref.~\cite{1997Law,1998Ao}.
Later on, with cold-atom mixtures, we can verify an original theoretical study on rotating properties of two
cold-atom species in Ref.~\cite{2003Kasamatsu}, which followed by suggestions of possible
realizations of ferro-fluidity with two-component dipolar systems~\cite{2009Saito}.
These investigations with binary atomic species became relevant as facing new challenges due to quantum degeneracy
for different kinds of atoms, including fermionic isotopes, mixtures of Bose condensates, superfluidity, and
Josephson tunneling, as pointed out in Ref.~\cite{2016Pitaevskii}.
Some experimental realizations with cold-atom mixtures, as the ones reported in Refs.~\cite{2014Barbut,2016Ulmanis},
have provided realistic basis for heteronuclear ultra-cold chemistry, which emerged as a new field of interest
with intense research activities in recent years (see, e.g., Refs.~\cite{2019Yang,2020Green,2020Shalchi} and quoted citations).

Relevant in these cases with binary systems are the miscible and immiscible properties,
which are derived from relations between the atoms inter- and intra-species two-body
interactions~\cite{1997Law,1998Ao}, and also controlled by the confinement~\cite{2012Wen,2017Bandyopadhyay}.
The control in experimental realizations with different atomic species
can be followed by the laboratory activities with ultra-cold molecular systems, as verified for example in
Refs.~\cite{2003Jochim,2008Thalhammer,2014Barbut,2016Ulmanis,2018Trautmann,2018Ilzhofer}.
Specifically, in Ref.~\cite{2008Thalhammer}, the dual species with $^{87}$Rb and $^{41}$K were confined in an optical
dipole trap in the proximity of inter-species Feshbach resonances. This system was also recently reported
in Ref.~\cite{burchianti2020} with attractive interactions.
There is also an increasing interest on investigating superfluid mixtures of condensates, which can be 
experimentally probed as reported in Ref.~\cite{richaud2019}.

Among the studies with dipolar bosonic quantum gases~\cite{2002Goral,2009Lahaye,2009Wilson,2015Bisset},
motivated by quantum ferrofluid instability observation~\cite{2015Kadau}, roton instability and droplet formation 
with dipole-dipole interaction were investigated in Refs.~\cite{2016Xi}, by solving cubic-quintic Gross-Pitaevskii (GP) formalism~\cite{2000Gammal,2001Abdullaev}, where it was pointed out the significant role of three-body 
interaction in the droplet formations. Later on, the miscibility properties of two-component BECs were investigated 
in a few works by some of us~\cite{2017KumarJPC,2017Kumar,2019KumarJPB}, using two different dipolar and non-dipolar
hyperfine spin states of a single isotope. These studies were followed by considering mass-imbalance
and rotating effects with different isotopes or atomic species in Ref.~\cite{2020Kumar}.
 Moreover, the studies with dipolar systems presented in Ref.~\cite{2019Kumar} are of particular interest  
 for possible experimental realizations, in which the dipolar interactions are shown to be
instrumental to control and tune the interactions of rotating binary mixtures, as well as for the spatial
separation of the species.
Within a Bogoliubov-de Gennes (BdG) calculation, by exploring possible photonic and rotonic phase
transitions, it was pointed out in Ref.~\cite{2021Lee} the relevance of the confinement geometry, in a work 
considering the miscibility and stability of dipolar bosonic mixtures.

The dynamics of binary condensates have also shown remarkable effects in rotating systems~\cite{2013ishino}. 
The activities in this direction have been intensified by investigations in which the stability of a
system can be probed by considering orbital angular momentum analysis.
Among other works, we can mention a miscibility analysis that was performed in Ref.~\cite{malomed2019},
for a binary condensed system in a ring geometry; in Ref.~\cite{2020nicolau}, the authors reported a study with repulsive
Bose-Einstein condensates (BECs) trapped in a two coupled rings configuration.
The Ref.~\cite{2020kanai} is another recently reported work considering processes in rotating BECs with angular quantum
momentum and torque transfer.
These kind of investigation with coupled systems are of interest due to the actual experimental possibilities with
tunable two-species coupled systems~\cite{2000trippenbach,2008papp,2018cabrera,2018Semeghini,2019errico},
which could be applied to condensates confined in spherical geometries by simulating micro-gravity conditions.

On the properties of coherent matter-wave bubbles, the interest started with the investigations reported
in Refs.~\cite{2001zobay,2004zobay}, in which one can find a proposal of possible experimental schemes.
Recent realistic possibilities in performing cold-atom experiments with low-gravity conditions at the international space
station (ISS)~\cite{NASA-BEC,2018elliott,2019lundblad} have drawn particular attention to the studies of condensates
confined in bubble traps, as exemplified by Refs.~\cite{tononi2019,2019prestipino,bereta2019, diniz2020, tononi2020,padavic2020}.
More recently, following observations aboard the international space station with  
ultracold atom bubbles created by using radio-frequency~\cite{2020aveline}, it was further reported
investigations on the nature and properties of bubble configurations with different sizes in Ref.~\cite{2021carollo}.
Among other future experiments in microgravity conditions, it was pointed out in this reference 
the real perspectives to generate vortices in condensate bubbles through distinct mechanisms, as 
direct stirring, trap rotation, or spontaneous generation across the condensate phase transition. In view of such 
advances in the control of condensed bubble generations at the ISS, it seems plausible to believe that
further experimental control can be reached in order to tune interacting binary condensed mixtures. 
With bubble-trapped condensates, we can also point out the recent studies performed in  
Refs.~\cite{2021bereta,2021tononi}, related to singly quantized vortices and superfluidity.

By considering repulsive Bose-Bose mixtures,  following a previous suggestion in Ref.~\cite{2015petrov},
it was recently predicted in Ref.~\cite{2021Naidon} a mixed-bubble regime in which bubbles of the mixed phase
coexist with a pure phase of one of the components.  Such an interesting study in which self-bound droplets
are stabilized by the repulsive Lee-Huang-Yang (LHY)~\cite{1957LHY} energy contribution, has been also 
verified in a recent
experimental realization, as reported in Ref.~\cite{2021skov}. This is a beyond-mean-field effect that occurs for
unequal masses or unequal intra-species coupling constants, being due to a competition between the mean-field
term, quadratic in densities, and a non-quadratic beyond-mean-field correction (For a related review, with
updated bibliography, see Ref.~\cite{2021luo}). However,  our following approach still relies on the mean-field
GP formalism. We do not consider here the possibility of canceling the
two-body interactions and consequent  overvaluing the LHY term correction, with the outcome of 
droplets formation.

Our motivation is concerned with the aforementioned theoretical and experimental
interest, in view of existing laboratory facilities to investigate ultra-cold atomic BEC systems confined in
circular and spherical geometries. By considering previous studies with single confined species, we concentrate
the present analysis in clarifying the dynamical behavior of binary atomic mixture, confined within the skin of a
three-dimensional (3D) spherical trap.
By assuming the initial stationary condition of the binary mixture as homogeneous and trapped at the surface
of a rigid sphere with radius $R$, the system is effectively two-dimensional (2D), with
all dynamics described by the two polar angles $\theta$ and $\phi$.
Such simple spherical geometry with two-species confined at a surface of a bubble, hopefully, can be useful to setup
initial experimental conditions for some related investigations, as well as for different other kinds of studies, such as
when assuming deformed radial geometries, or by considering atomic species with more involved inter- and
intra-species interactions, as the case of dipolar binary systems in spherical geometries.

For the next, this paper is structured as follows:  The basic framework of the mean-field model formalism is
introduced in Sec.~\ref{sec2}.
In Sec.~\ref{sec3}, our approach in probing the stability of an original stationary solution is exemplified by
applying the method to the non-vorticity case of a homogeneous two-species mixture.
In Sec.~\ref{sec4}, we study the stability of stationary states with quantized vortices in the shell, by considering
the specific case with both species having opposite charge vorticity, $s_2=-s_1=1$.
In this case, a variational approach is shown to be helpful in establishing analytical solutions in support to
the full-numerical ones.
The dynamics and stability of the states are studied in detail in Sec.~\ref{sec5}, with analysis of time-evolution
of the unstable modes.
Finally, in Sec.~\ref{sec6}, we present our conclusions and outlook. An appendix is also included concerning our
numerical method for real-time integration of the GP equation.

\section{Model formalism}\label{sec2}

In our present study, we assume two  atomic species ($i=1,2$) with the same mass $M$, which are
initially within a homogeneous mixture, with both species having the same density. 
Apart from the theoretical convenience to consider a more symmetric initial configuration, 
which also will facilitate the analysis of the expected pattern results, studies with equal-mass binary 
systems are supported by existing BEC experiments with two-spinor states of the same 
isotope~\cite{1996Ho,1997Myatt}. 
The stability of the initial configuration will be studied by adding a small time-dependent perturbation 
in the initial configuration. For the stability, we observe that our study is concerned only with the 
occurrence of {\it dynamical instability} in the system.
The possibility of {\it energetic instability} of the coupled condensates 
 is not being considered in the present work, in which we have assumed zero temperature $T$.
 Energetic instabilities play role only in excited states, as they need a way to get rid of energy, which in 
 BEC systems can be accomplished by losing  the extra energy through contact with the thermal 
 cloud~\cite{2016Pitaevskii}.
 They could occur in our approach only for $T>0$, when having excited angular momentum 
 states, with vortices in a spherical geometry~\cite{2009jackson,padavic2020}.
In that case, the generated vortices shall migrate to the equator-line of the sphere and being annihilated.
However, for that, a dissipation mechanism is required, as interactions between the condensate with the
thermal cloud. Here, we are concerned with Bose gases at effectively zero temperature;
thus, in practice, there is no thermal cloud to allow a dissipation mechanism.

The inter- and intra-species interactions are given by $g_{ij}\equiv (4\pi\hbar^2 a_{ij}/M)$, where $a_{ij}$
is the atom-atom scattering length.
 As the interaction ratio between inter-species and intra-species, $g_{12}/g_{ii}$ (by assuming
 $g_{11} = g_{22}$) increases,
 unstable modes causing inhomogeneities should appear, analogous to the one-dimensional
 (1D) ring case~\cite{malomed2019}. Therefore, in our following approach, we assume the
 two atomic species are confined in the skin of a 3D spherical shell with fixed radius $R$, implying on the
 existence of the condensate densities only inside an infinitesimal range, with $R(1\pm\delta/2)$.
This trapped region covering the whole sphere we can label as $V,$ the total volume of the confinement,
given by $V\sim{4\pi}R^3\delta$.
By modeling the system with an effective $\delta\approx 0$,
we can write the density states for each component $i$, with the radial function given in terms of the
Dirac delta $\delta(r-R)$, such that  $|\Psi_i({\bf r},t)|^2 =\delta(r-R)|\psi_i(\Omega,t)/R|^2$, where 
$\Omega\equiv(\theta,\phi)$ gives the angular position in the sphere.

 In order to validate approximately the assumption $\delta\approx 0$, we need to estimate the
 level energies of a trapping interaction in the radial direction, for an infinite potential well, centered in R, 
 with radial size $R\delta/2$ [radial wave function being zero at $R(1\pm\delta/2)$]. 
 For a $\delta$ enough small, the energy difference between the ground 
 and first excited state should be enough large.  For such an estimate, we can follow section VI and VII of 
 Ref.~\cite{bereta2019}, in which they consider a thicker shell for the confining region of a given condensate.
 By following this approach, we can verify that the single-particle radial energy excitation (ground to first 
 excited states), in energy units $\hbar^2/(MR^2)$, is given by ${\cal E}_R=3\pi^2/(2\delta^2)$.
 Correspondingly, in the same units, the estimated absolute value of the energy obtained from the non-linear
 quartic interaction term, for each species $i$ with scattering lengths 
 $a_{ii}$, with the condensed particles in the ground-state level, is given by  
 ${\cal I} = 3N_i |a_{ii}| /(4R \delta)$.
 Therefore, for ${\cal I}\ll{\cal E}_R$, we need $\delta \ll 2\pi^2 R/(N_i |a_{ii}|)$.
 This indicates the strict range of validity of our reduction from 3D to the hard 2D sphere, which can be 
accomplished by controlling the two-body scattering lengths and the number of atoms. 
The two-body inter- and intra-species interactions can be written as dimensionless parameters by 
$\gamma_{ij}\equiv g_{ij}N_j/{R^3}=4\pi N_j(a_{ij}/R)$,
in which it was included the density dimension ($1/R^3$) and the number of atoms $N_{j}$.
In our case, the above estimative for the 2D reduction is given by $\delta\ll  8\pi^3/|\gamma_{ij}|$.
In order to bring this estimate to realistic values of the physics parameters,  we first note that
the two-body interactions $a_{ij}$ can be tuned by using Feshbach resonance mechanisms, 
with its absolute value varying from almost zero to very large values as $100 a_0$, where $a_0$ 
is the Bohr radius.  
On the bubble dimensions, according to Ref.~\cite{2021carollo}, the radial sizes $R$ can be of 
the order of 1 mm or even larger,  with the bubble thickness $R \delta$ being of the order of 
few $\mu$m. 
For instance, let us assume $|a_{ij}|\sim 100 a_0$, with $R\sim 100 \mu$m $\approx 
2\times 10^6 a_0$. In this case, $|\gamma_{ij}|=4\pi N_j |a_{ij}|/R\sim 2\pi N_j \times 10^{-4}$, 
which should be  within the covered range of $|\gamma_{ij}|$ values to be considered for the 
strict validity of the 2D reduction that we are assuming. As realistic values for the thickness are 
of the order of few $\mu$m, we can take  $\delta\sim 10^{-2}\to 10^{-3}$. 
So, a 3D treatment may be required only when very large values of $\gamma_{ij}$ are 
considered. In such a case, our 2D approach is expected to provide a good approximation.

By assuming $R$ to be our length scale, with the time unit given by $M R^2/ \hbar$, 
with the states $\psi_{i=1,2}\equiv \psi_i(\Omega,t)$ normalized to one,
the original nonlinear Schr\"odinger equation is reduced to the following dimensionless coupled
equation:
 \begin{eqnarray}
{\rm i}\frac{\partial \psi_i}{\partial t}
&=&\left[\frac{1}{2} {\bf L}^2 +
\sum_{k=1,2}\left(\gamma_{ik}|\psi_k|^2\right)\right]\psi_i,
\label{eq01}
\end {eqnarray}
where ${\bf L}\equiv -{\rm i}\left[\hat{e}_\phi\frac{\partial}{\partial\theta}
-\hat{e}_\theta\frac{1}{\sin\theta}\frac{\partial}{\partial\phi}\right]$
is the dimensionless angular momentum operator ($\hat e_\phi$ and $\hat e_\theta$ being,
respectively, unit vectors along the azimuthal $\phi$ and polar $\theta$ directions), with
{\small\begin{eqnarray}
{\bf L}^2&\equiv&-\left[\frac{1}{\sin\theta} \frac{\partial}{\partial \theta}
\left(\sin\theta\frac{\partial}{\partial \theta}\right)
-\frac{1}{\sin^2\theta} L_z^2
\right],
\label{eq02}
\end{eqnarray}
}where $L_z^2\equiv -\frac{\partial^2}{\partial\phi^2}$.
With the total two-component wave-function $\Psi(\Omega,t)\equiv \Psi[\psi_1,\psi_2]$ normalized
to the number of atoms $N$, such that
{\small$\int d\Omega |\Psi(\Omega,t)|^2=$ $N_1+N_2\equiv N$},
the functional energy corresponding to (\ref{eq01}) is given by
{\small
\begin{eqnarray}
    E[\psi_1, \psi_2]    =
    \sum_{i}\frac{N_i}{N}
    \int \!\! \mathrm{d}\Omega \left[
     \frac{\left|{\bf L}\psi_i\right|^2}
    {2} +\sum_{j}\frac{\gamma_{ij}}{2}|\psi_{i}|^2 |\psi_{j}|^2
    \right].
\label{eq03}\end{eqnarray}
}The inter- and intra-species two-body interactions, respectively  $\gamma_{ij}$ and $\gamma_{ii}$, in
our approach, are assumed that in general can be repulsive ($>0$) or attractive ($<0$), with
possible static and dynamical solutions being considered.
Let us consider initially the non-perturbed stationary solutions of Eq.~\eqref{eq01}, normalized to
one, with chemical potentials $\mu_i$, as given by
\begin{equation}
    \psi_{i0}\equiv\psi_{i0}(\theta, \phi,t) = \frac{f_i(\theta)}{\sqrt{2 \pi}} e^{{\rm i} (s_i \phi-\mu_i t)},
    \label{eq04}
\end{equation}
in which the  $\phi$ dependences are assumed with given charges $s_i \in \mathbb{Z}$,
defined as the initial vorticity of the components $i$, which yields
{\small\begin{equation}
\langle L_z\rangle_i= -{\rm i}\int d\Omega\psi_i^*\frac{\partial\psi_i}{\partial \phi} =
s_i \int_0^\pi d\theta \sin\theta |f_i(\theta)|^2=s_i
,\label{eq05}\end{equation}
}such that when non-zero can create vortices on the sphere. It is worth to stress that $s_i\neq 0$
affects the $f_i(\theta)$ boundary condition, due to the $1/\sin^2{\theta}$ Laplacian term,
implying in $f_i(\pi) = f_i(0) = 0$ for $s_i \neq 0$.

The stationary solutions of the linear part of Eq.~(\ref{eq01}) (when $\gamma_{ik}=0$) are the well-known
spherical harmonics $Y_{\ell_i,s_i}(\theta,\phi)$ ($\ell_i=0\to\infty$, $-\ell_i\le s_i\le \ell_i$) in which the associated 
Legendre functions $P_{\ell_i,s_i}(\theta)$ are the solutions in the $\theta$ variable.
Therefore, as considering the interactions, with  $\gamma_{ik}\ne 0$, we can generate a multiplicity of stationary
solutions for a coupled system, which emerge from the linear ones. These nonlinear solutions can be continued
by increasing the nonlinearity of the system.

Within our aim to follow the simplest solutions and learn about their stability, as well as
the vorticity of these states when considering opposite values of the azimuthal quantum number
$s_i$ (which can be of interest in experimental setups), we start by considering the non-vorticity case,
in which $s_i=0$. For that, we first notice that the spherical harmonic state with $\ell_i = 0$, simply given by 
the constant $1 / \sqrt{4 \pi}$, is the ground state of the linear part and also a stationary state of the 
full nonlinear problem.  Although, it is not necessarily the ground state of the mixture, as we should 
consider the miscibility of the coupled system through the intra- and inter-species interactions.
In this regard, we can split the possible non-linear stationary solutions of \eqref{eq03} into two distinct
cases, related to the miscibility of the mixture. More precisely, by following a simple energetic consideration
as given in Ref.~\cite{1998Ao},  comparing the non-linear energy contribution of a complete miscible 
configuration, with a complete immiscible one, in which the two species are not interacting (but constrained 
within the same total volume), 
one can show that the immiscible configuration ($g_{12}>\sqrt{g_{11}g_{22}}$) provides the lower energy.
However,  this analysis is not providing an exact relation at which the system becomes unstable. 
For that, a more detailed stability analysis is required, in which the trap geometry and kinetic energy 
term can also be relevant. This will be shown in the present case that we have the confinement region 
on a spherical surface.

In Sec.~\ref{sec3} we explore the stability of the uniform state to exemplify our scheme in a full 
analytical case without vorticity. In Sec.~\ref{sec4}, the approach is applied to the hidden vorticity case,  
where we consider the $\ell_i=1$ solution for the linear part, with $s_i=\pm 1$.

The stability of these states is studied by assuming they are submitted to infinitesimal time-dependent
perturbations $u_{i\nu}\equiv u_{i\nu}(\theta,\phi)$ and $v_{i\nu}\equiv v_{i\nu}(\theta,\phi)$,
with oscillating modes $\omega_\nu$, with perturbed solutions given by
{\small\begin{eqnarray}
\psi_i(\theta,\phi,t)=  \psi_{i0}+\left[u_{i\nu} e^{-{\rm i}\omega_\nu t} +v_{i\nu}^*e^{{\rm i}\omega_\nu^*t} \right]
e^{{\rm i}(s_i\phi-\mu_i t) }.
\label{eq06}
\end{eqnarray}
}By considering a linear stability analysis, $u_{i\nu}$ and $v_{i\nu}$ will also be associated
to integer quantum numbers $\nu$, which are representing the perturbation modes being considered
as superposition to the stationary states.

\section{Homogeneous non-vorticity case}\label{sec3}

Let us first consider the Bogoliubov modes on top of  homogeneous states with no vorticity,
by following Refs.~\cite{malomed2019, kreibich2012}, such that $s_i = 0$.
By looking at the solutions
of the linear part of Eq.~\eqref{eq01}, we noticed that they are given by the usual Legendre polynomials
$P_{\ell_0}(\cos\theta)$ that are eigenfunctions of ${\bf L}^2$ with eigenvalues $\ell_0(\ell_0+1)$, with
$\langle L_z\rangle_i=0$.
As instructive in the present section, for the stationary solutions we consider the simplest
ground-state $\ell_0=0$, such that  $f_i(\theta) = 1/\sqrt{2}$ in Eq.~\eqref{eq04}, which also
reduces the non-linear term to a constant, given by the binary interactions.
Next, in order to probe the stability of these solutions under small perturbations,
in our analytical approach we assume the perturbations are given by eigenfunctions of the angular momentum operator 
${\bf L}^2$ and $L_z$; namely, the spherical harmonics,
$Y_{\ell,m}\equiv  Y_{\ell,m}(\theta,\phi)$, with $(\ell,m)$ being the corresponding quantum numbers.
Therefore, with the full angular dependence of the perturbation expanded in spherical harmonics,
considering in Eq.~\eqref{eq06} $s_i=0$, with $\nu\equiv\ell$ and the frequency oscillating modes
given by $\omega_\nu=\omega_\ell$, we have
\begin{eqnarray}
u_{i\nu}(\theta,\phi)\equiv u_{i\ell}  Y_{\ell,m},\;\;
v_{i\nu}(\theta,\phi)\equiv v_{i\ell}  Y_{\ell,m}.
\label{eq07}
\end{eqnarray}
 The oscillating modes are asumed in general as complex quantities,
$\omega_\ell \equiv {\cal R}e(\omega_{\ell})+{\rm i}{\rm Im}(\omega_{\ell})$, such that stable solutions imply ${\rm Im}(\omega_{\ell})=0$ for all possible values of $\ell$.
By replacing (\ref{eq06}) with (\ref{eq07}) in (\ref{eq01}), followed by a linearization, which retains
only the first order terms of $u_{i\ell}$ and $v_{i\ell}$ in the nonlinear part, we obtain the respective
BdG coupled equations~\cite{brtka2010}. Defining $\tilde{\ell}^2\equiv \ell(\ell+1)$ to simplify
the following formalism, from Eqs.\eqref{eq01}, \eqref{eq06} and \eqref{eq07}, we obtain
{\small
\begin{eqnarray}
&\mu_i&=\sum_{k=1,2} \frac{\gamma_{ik}}{4\pi},
\nonumber \\
&0&=\left(\omega_{\ell}-\frac{\tilde{\ell}^2}{2}\right)u_{i\ell} e^{-{\rm i}\omega_\ell t}Y_{\ell,m}
-\left(\omega_\ell^*+\frac{\tilde{\ell}^2}{2}\right) v_{i\ell}^* e^{{\rm i} \omega_\ell^*t} Y^{*}_{\ell,m}\nonumber\\
&-&\sum_k \frac{\gamma_{ik}}{4\pi} \bigg\{\left(u_{k\ell} +v_{k\ell}\right)e^{-{\rm i}\omega_\ell t} Y_{\ell,m}
+ c.c. \bigg\},\label{eq08}
\end{eqnarray}
}in which the expression for $\mu_i$ could be directly verified from Eq.~(\ref{eq01}), considering
the normalization of the stationary solutions.
The linear independence of  $e^{-{\rm i}\omega_\ell t}$ and $e^{{\rm i}\omega_\ell^{*} t}$ in
(\ref{eq08}) implies in two separate equations, leading to a relation between $u_{i\ell}$ and $v_{i\ell}$,
{\small\begin{eqnarray}
\left(\omega_\ell-\frac{\tilde{\ell}^2}{2}\right)u_{i\ell}=
-\left(\omega_\ell+\frac{\tilde{\ell}^2}{2}\right) {v}_{i\ell} =
\sum_{k=1,2}\frac{\gamma_{ik}}{4\pi}\left({u}_{k\ell}+{v}_{k\ell} \right),
\nonumber\end{eqnarray}
}with the solution for the oscillating modes given by
{\small\begin{eqnarray}\hspace{-0.5cm}
\omega_{\ell,\pm}^2&=&\frac{\tilde{\ell}^2}{2}\left[\frac{\tilde{\ell}^2}{2}+
\frac{\gamma_{11}+\gamma_{22}
\pm\sqrt{(\gamma_{11}-\gamma_{22})^2 +4\gamma_{12}\gamma_{21}}}{4\pi}
\right]\hspace{-1mm}.
\label{eq09}
\end{eqnarray}
}As a general outcome from the above, we note that the solutions become unstable
when assuming overall attractive interactions, such that the second term within the square
brackets is negative, with absolute value larger than $\ell(\ell+1)/2$.
In our following approach along this work, we are assuming that $\gamma_{12}=\gamma_{21}$,
implying that the particle numbers are the same for both species ($N_1=N_2$), considering
that $g_{12}=g_{21}$ and both particles have the same mass. For the intra-species interactions, 
we are assuming $\gamma_{11}=\gamma_{22}$, which can be easily satisfied
by altering the two-body scattering lengths using Feshbach resonance techniques~\cite{Timmermans1999}.
Therefore, we obtain
\begin{eqnarray}
\omega^2_{\ell,\pm} = \frac{\ell(\ell+1)}{2}\left[\frac{\ell(\ell+1)}{2}+
\frac{\gamma_{11}\pm|\gamma_{12}|}{2\pi}\right].
\label{eq10}
\end{eqnarray}
In this case, the inter-species interaction $\gamma_{12}$ being attractive or
repulsive is not relevant, as the results related to stability should be the same.
 The fact that the results do not depend on the sign of $\gamma_{12}$ is related
to the simple homogeneous spherical symmetry we are considering.
Besides that, for $\gamma_{11}>0$ only the minus sign branch can be unstable subject 
to the condition $|\gamma_{12}| > \gamma_{11} + \pi \ell(\ell + 1)$.
Another point is that, for $\gamma_{11}<0$, we can only have a small stability branch if
the kinetic energy term $\ell(\ell+1)/2$ is dominating the term inside the square brackets
of \eqref{eq10}.

Another simple possibility occurs for $\gamma_{11}=-\gamma_{22}$ in Eq.~\eqref{eq09},
which will result that the oscillating modes become independent of the signs of both
intra- and inter-species interactions ($\gamma_{11}$ and $\gamma_{12}$) given by
{\small\begin{eqnarray}\hspace{-0.5cm}
\omega_{\ell,\pm}^2&=&\frac{\ell(\ell+1)}{2}\left[\frac{\ell(\ell+1)}{2}\pm
\frac{\sqrt{\gamma_{11}^2 +\gamma_{12}^2}}{2\pi}
\right]\hspace{-1mm}.
\label{eq11}
\end{eqnarray}
}In this case, the stability frontiers for each perturbation mode $\ell$ are circles 
with radius $\pi \ell (\ell + 1)$, due to the square brackets term.

\begin{figure}[h]
    \centering
    \includegraphics[scale=0.23]{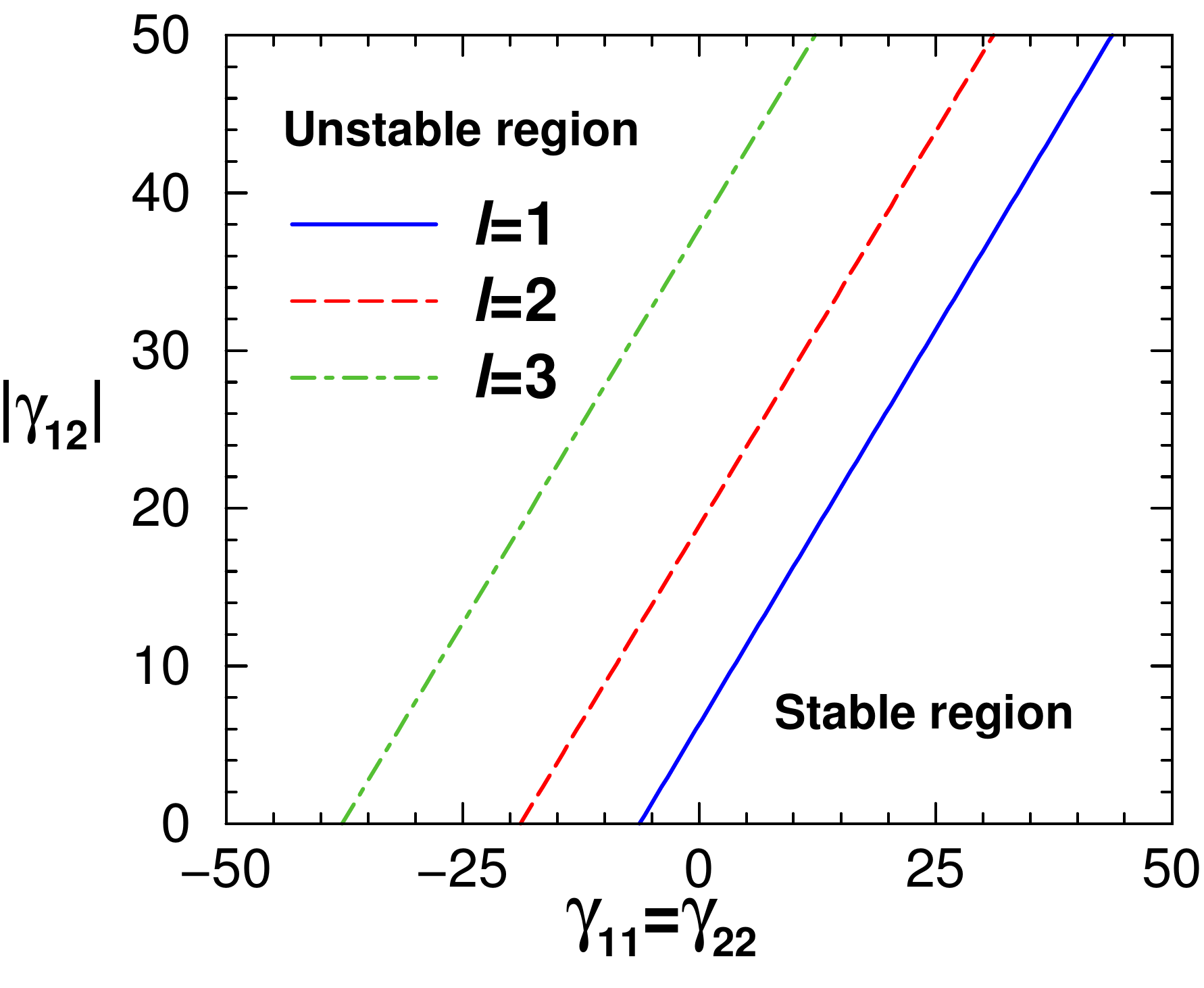}
    \includegraphics[scale=0.23]{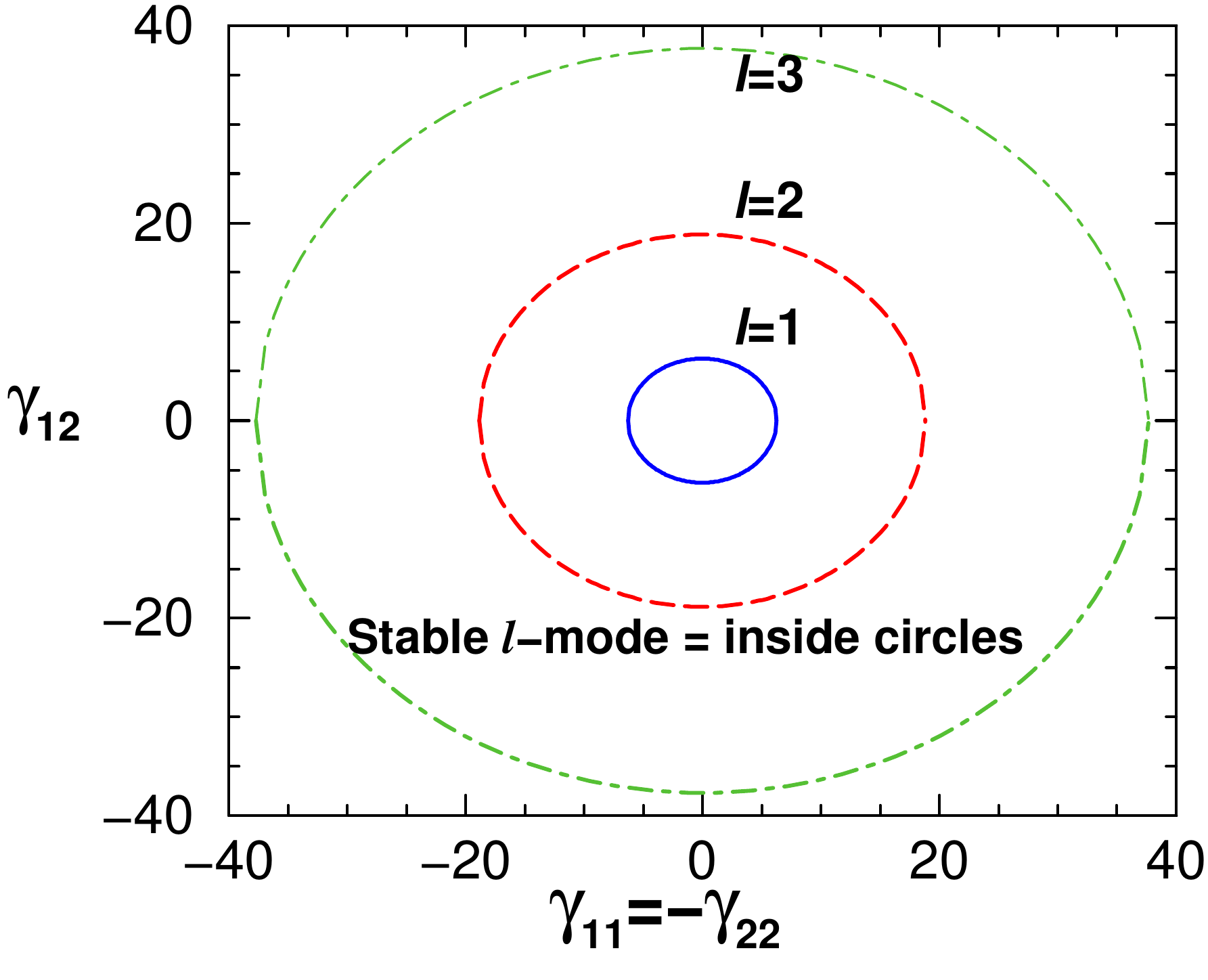}
\vspace{-0.2cm}
\caption{(Color on-line)
The stable and unstable regions are represented, for $\gamma_{12}$ versus
$\gamma_{11}$, indicating the miscible and immiscible phase regions respectively.
In (a), for $\gamma_{11}=\gamma_{22}$, the stable regions are right-below the line 
modes with $\ell=$1, 2, and 3; whereas, for $\gamma_{11}=-\gamma_{22}$ (b),
they are inside circles.  All quantities are dimensionless, with units defined in the text.}
\label{fig-01}
\end{figure}

In both cases, given by Eqs.~(\ref{eq10}) and (\ref{eq11}), the stable and unstable regions
are represented by diagrams of $\gamma_{12}$ versus $\gamma_{11}$ in the two panels of
Fig.~\ref{fig-01}, in which the miscible phases are identified as stable regions.

The present homogeneous case, with the stationary solution in the ground state, is quite
simple, as verified by the corresponding chemical potential given by Eq.~(\ref{eq08}), such that
it serves the purpose to clarify the approach we are going to consider for the non-homogeneous
case, with vorticity (in which we assume $s_i\ne 0$).
Within both conditions, (a) $\gamma_{11}=\gamma_{22}>0$, given by Eq.~(\ref{eq10}),
and (b) $\gamma_{11}=-\gamma_{22}$, given by Eq.~(\ref{eq11}),  the instabilities are
prescribed by  the threshold for the imaginary frequency modes, which are, respectively, given by
{\small
\begin{eqnarray}
{\rm Im}(\omega_{\ell,-}) =
\left\{ \begin{array}{l}
\pm\left[\frac{\tilde{\ell}^2}{4\pi}
\left(|\gamma_{12}|-\gamma_{11} -\pi\tilde{\ell}^2\right)\right]^{\frac12}
\bigg|_{\gamma_{11}=\gamma_{22}},
\\
\\
\pm\left[\frac{\tilde{\ell}^2}{4\pi}\left(\sqrt{\gamma_{11}^2+
\gamma_{12}^2}-\pi\tilde{\ell}^2\right)\right]^{\frac12}
\bigg|_{\gamma_{11}=-\gamma_{22}}.
\label{eq12}
\end{array}\right.
\end{eqnarray}
}However, we should notice that the second case ($\gamma_{11}=-\gamma_{22}$) is already
contained in the first case, given the following replacement:
$\gamma_{11}\to 0$ and $\gamma_{12}\to\sqrt{\gamma_{11}^2+\gamma_{12}^2}$.
From Eqs.~(\ref{eq09})-(\ref{eq12}), we can also verify that, given the non-linear interaction term,
there is a critical upper value $\ell=\ell_{max}$, which contributes to ${\rm Im}(\omega_{\ell,-})$.
However, it should also be clear that the lower level modes with $\ell<\ell_{max}$ are already
establishing the instability of the system.

For simplicity, in the following we select the condition $\gamma_{11}=\gamma_{22} = 10$, which is
given by Eq.~(\ref{eq10}), for comparison of the analytical with full-numerical results. In contrast
to the analytical results where we can explicitly select the angular momentum quantum numbers, in
the numerical approach we start from the general perturbation form in Eq.~\eqref{eq06} and only
factor out the azimuthal exponential part of the spherical harmonics.
This procedure then demands a discretization of the differential
operator $\textbf{L}^2$ in $\theta$ coordinate, which provides a general platform to also handle non-uniform states, 
as will be needed in the next section to study vortices. Consequently, within a numerical approach
we have an arbitrary indexing $\nu$ of the states according to Eq.~\eqref{eq06}, which is fixed
provided that the imaginary part of the eigenvalues $\omega_{\nu}$ are in decreasing order. More details
of the BdG system in the numerical approach are provided in the next section.

\begin{figure}[h]
\centering
\includegraphics[scale=0.26]{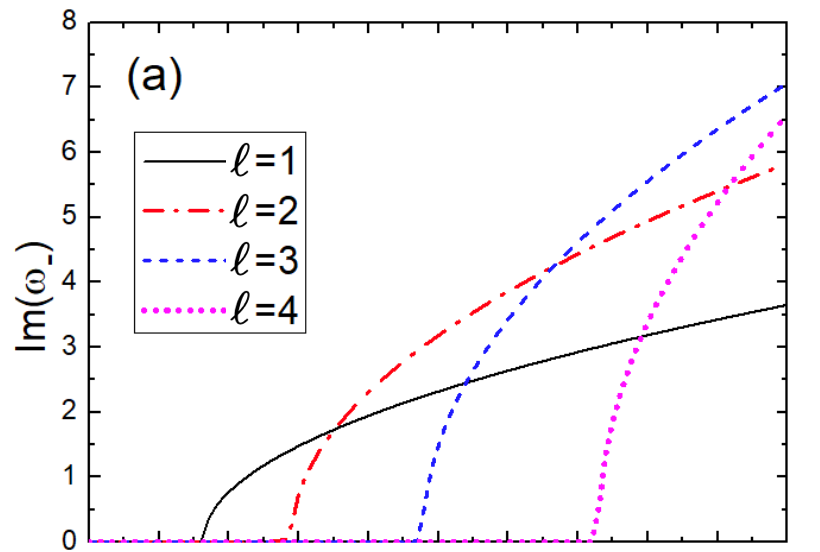}
\includegraphics[scale=0.26]{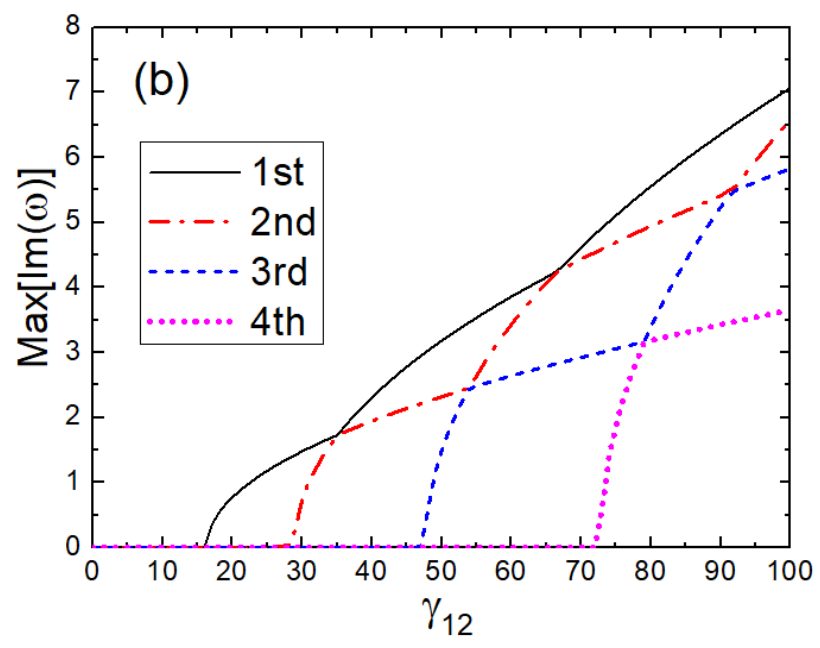}
\includegraphics[scale=1.04]{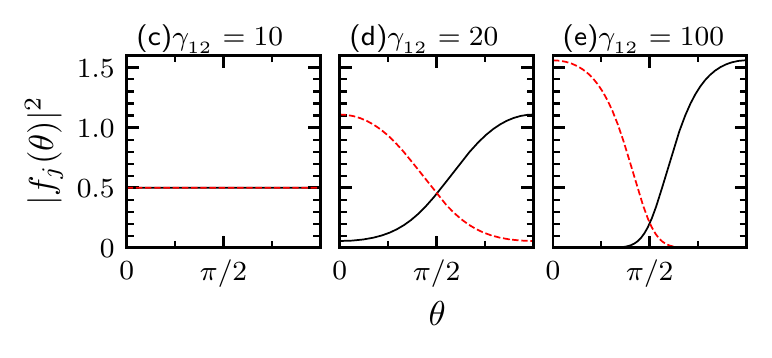}
\vspace{-.5cm}
\caption{\label{fig-02}
(Color on-line) In frame (a), the imaginary frequency mode values [positive branch of
${\rm Im}(\omega_{\ell,-})$, for $\ell=$1, 2, 3, and 4, given by Eq.~(\ref{eq12})],
are shown as functions of the inter-species interaction $\gamma_{12}$, by considering
$\gamma_{11}=\gamma_{22}=10$.
In frame (b), the numerical solutions for the maximum values of ${\rm Im}(\omega_{\nu})$,
given by \eqref{eq06}, shows the exact correspondence with frame (a), with
the legend indicates the order of dominance of the modes.
Correspondingly, the precise ground-state densities $|f_j(\theta)|^2$ ($\theta$ in rad) 
for both species $j=1$ (dashed lines) and $j=2$ (solid lines), are shown in 
(c), (d) and (e), for three specific values of $\gamma_{12}$ (indicated at the top), 
without constraining the solutions to a uniform miscible state.
Concomitantly to the appearance of the instabilities of miscible uniform state, the actual 
ground state enters an immiscible phase. All quantities are dimensionless, with units 
defined in the text.
}
\end{figure}

In Fig.~\ref{fig-02}, our results are shown for the imaginary spectrum, in which the different
instability modes are being explicitly identified. In panel (a), Eq.~\eqref{eq10} is used and the
modes are indexed by their angular momentum quantum number. In panel (b), the numerical results
are presented using numerical diagonalization and the modes are sorted in decreasing order. The
critical values for $\gamma_{12}$ at which the instabilities
start are given by $|\gamma_{12}|_{crit}\simeq$ 16.28, 28.85, 47.70, and 72.84 for the
unstable modes with $\ell=$1, 2, 3, and  4, respectively.
Clearly, this shows that the geometry of the system can extend the stability criterium 
beyond the $g_{12}> \sqrt{g_{11}g_{22}}$,  as also verified in the case of ring 
geometry~\cite{malomed2019}.

Panels (a) and (b) of Fig.~\ref{fig-02}, with the analytical and numerical results respectively, show a notable
agreement of both approaches. Nevertheless, in the numerical approach we cannot separate explicitly the
total angular momentum numbers $\ell$, since in this case no constraints are implied a priori for $\theta$ 
coordinate, providing us a general method applicable even for non-uniform stationary states. Instead, 
we can only sort all imaginary eigenvalues in ascending order, as mentioned above. Despite this caveat, a 
correspondence is verified between the numerical levels of instabilities with the analytical $\ell-$mode 
solutions, such that the overall results are identical.

In Fig.~\ref{fig-02}, we also add three ground-state density plots as function of $\theta$ in panels (c), (d), 
and (e), which are obtained numerically, without constraining the species to be completely miscible. 
Therefore, as the instability modes grow from zero, the lowest energy state enters an immiscible phase, 
and the overlap becomes smaller as larger is the inter-species coupling $\gamma_{12}$.

The analysis of these results, obtained in a simple no-vorticity full-analytical situation,
is instructive to guide us in the analysis of the instabilities that occur in vortex states.

\section{Quantized vortices on a bubble}
\label{sec4}
In our approach, we are assuming that both species are initially with the same density, in the lowest 
non-interacting stationary states $\ell_i=1$, and with opposite charge vorticity between the 
components, given by $s_2=-s_1=1$ in Eq.~\eqref{eq04}~\cite{2016Pitaevskii}.
Therefore, we are considering initially the states within a complete miscible configuration and hidden-vorticity,
whereas by hidden we mean that there is no net angular azimuthal momentum despite each species have a 
single charged vortex.
As mentioned right after Eq.~\eqref{eq04}, this choice $s_{1,2} = \pm 1$ implies that $f_j(\theta)$ must vanish
at the poles to avoid divergences in the kinetic energy. Therefore, homogeneous states are no longer allowed
in such case of hidden vorticity (HV), which will bring more restrictions on the possibility of 
analytical solutions for the stationary equations and for the BdG stability analysis,
when considering nonzero interaction.

From Eqs.~\eqref{eq01}, \eqref{eq02}, and \eqref{eq04}, 
we obtain the corresponding stationary eigenvalue equation, with $f_i(\theta)$ being the eigenfunctions and 
$\mu_i$ the eigenvalues, given by:
{\small
\begin{eqnarray}
\mu_i f_i
&=& \left(\frac{1}{2}L^2_{s_i} +\sum_{k=1,2}\frac{\gamma_{ik}}{2\pi}f_k^2\right)f_i \label{eq13}\\
&=&-\frac{1}{2}\frac{d^2 f_i}{d\theta^2}-\frac{\cot\theta}{2} \frac{d f_i}{d\theta}
+\frac{s_i^2 f_i}{2\sin^2\theta} +\sum_{k=1,2}\frac{\gamma_{ik}}{2\pi}f_k^2f_i\nonumber
,\end{eqnarray}
}in which we are defining the dimensionless operator $L^2_{s_i}$ as the squared angular momentum operator
${\bf L}^2$ given in Eq.~\eqref{eq02}, after replacing the $L_z^2-$operator in favor of the
corresponding azimuthal quantum number, which is $s_i^2$ in this case.
For general solutions of the above non-perturbed stationary equation, as considering different possible
interactions $\gamma_{ij}$, which appear in the non-linear coupling term, we found appropriate to apply
numerical techniques, particularly by taking into account the stability analysis which will be followed
with time-dependent small perturbations.  As the corresponding linear counterpart of Eq.~\eqref{eq13}
has analytical solutions which are given by the associated Legendre functions, 
the numerical solutions are obtained starting with an analytical continuation from non-interacting case 
to arbitrary interacting parameters $\gamma_{ij}$.

Another relevant aspect is that in our approach the initial condition is given by a complete mixed configuration
of the two species. By considering that, in the following, we introduce a variational solution analysis 
for HV states, which will be compared with the corresponding full-numerical results.

\subsection{Variational treatment - homogeneous case}
For a variational solution, we consider here the vorticity case with $s_1=-s_2=1$, for which the lowest level
provided by the linear solution is given by $\ell=1$. Therefore, for such homogeneous case, 
we assume the corresponding linear solutions $f_i = \sqrt{3/4}\sin\theta$ being identical for both 
species and modified by a variational parameter $\beta$, such that $f_{v}^2 = \lambda(\sin\theta)^\beta$, 
with ${\lambda}$ given by the normalization of $f_v$.
This variational solution is obviously limited to $\beta> 0$, in order to have a normalized wave function.
With the above assumptions, the non-linear inter- and intra-species interaction parameters can 
be replaced by a single parameter $\gamma\equiv \gamma_{11}+\gamma_{12}$.
The component wave functions are given by
 $\psi_v\equiv\psi_1=\psi_2= \sqrt{1/(2\pi)} f_v(\theta)$, where $f_v\equiv
 f_v(\theta) = \sqrt{\lambda (\sin\theta)^\beta}$, are normalized such that
 \begin{eqnarray}
\int d\Omega |\psi_v|^2 = \lambda \int_0^\pi d\theta(\sin\theta)^{\beta+1}= \lambda{\cal J}(\beta)=1,
\label{eq14}\end{eqnarray}
from where we have an integral definition for ${\cal J}(\beta)$,
with the following properties ($\beta>0$):
\begin{eqnarray}
{{\cal J}(\beta)}&=&\frac{\beta}{\beta+1} {{\cal J}(\beta-2)},\nonumber\\
\frac{d{\cal J}(\beta)}{d\beta}&=&
 \int_0^\pi d\theta(\sin\theta)^{\beta+1}\ln(\sin\theta) .
\label{eq15}\end{eqnarray}
 With the above, the total energy (\ref{eq03}) can be written as a function of $\gamma$ and
 $\beta$, 
\begin{eqnarray}
E(\gamma,\beta)&=&\frac{(\beta+2)^2}{8\beta}+
\frac{\gamma}{4\pi}\frac{{\cal J}(2\beta)}{\left[{\cal J}(\beta)\right]^2}\nonumber\\
&=&\frac{(\beta+2)^2}{8\beta}+\frac{\gamma}{8\pi}
\frac{(\beta+1)^2}{\Gamma(2\beta+2)}\left[
\frac{\Gamma(\beta+1)}{\Gamma(\frac{\beta}{2}+1)}\right]^4
,\label{eq16}\end{eqnarray}
where the nonlinear term  was expressed in terms of the well-known gamma functions, 
$\Gamma(\zeta)$ [which is an extension of integer number factorial $\Gamma(n)=(n-1)!$].
Therefore, by minimizing $E(\gamma,\beta)$ for the parameter $\beta$, we obtain 
a relation between $\gamma$ and the variational $\beta$, such that we have also 
the corresponding minimized energies, given as $E_{var}(\beta)$.
The results of this variational procedure are displayed in the Fig.~\ref{fig-03}. 
In the left panel (a) we show the functional relation between the variational parameter 
$\beta$ and the interaction parameter $\gamma=\gamma_{11}+\gamma_{12}$; and, 
in the right panel (b) we present the variational results for the energy and chemical 
potential as functions of $\gamma$, together with the exact numerical results. 
As verified the variational results are providing the {\it almost exact} solutions, even 
for very large non-linearities.
The chemical potential, also shown, can be obtained directly from $E_{var}(\beta)$, 
\begin{eqnarray}
\mu_{var}(\beta) = 2 E_{var}(\beta) -  \frac{(\beta+2)^2}{8\beta}.
\label{eq17}
\end{eqnarray}
In case of attractive overall interactions ($\gamma_{11}+\gamma_{12}<0$) we observe 
in the left panel of
Fig.~\ref{fig-03} that the minimization of the energy will correspond to increasing values 
of $\beta$. In order to clarify the limit for large negative $\gamma$, we can apply in 
Eq.~(\ref{eq16}) the well-known Stirling's formula, derived for real positive variables $z\gg 1$
(shown to be valid even for relatively low values of $z$~\cite{stirling}), given by
$\Gamma (z+1) \sim \sqrt{2\pi z} \left({z}/{e}\right)^z.$
With this expression, for $\beta\gg1$, 
{\footnotesize\begin{eqnarray}&&\frac{(\beta+1)^2}{\Gamma(2\beta+2)}\left[
\frac{\Gamma(\beta+1)}{\Gamma({\beta}/{2}+1)}\right]^4=
e \sqrt{\frac{1}{2\pi}}\frac{(2\beta+2)^2(2\beta)^{(2\beta)}}{(2\beta+1)^{(2\beta+3/2)}}\nonumber\\
&&\sim e \sqrt{\frac{\beta}{\pi}}
\left(\frac{2\beta}{2\beta+1}\right)^{2\beta} = \sqrt{\frac{\beta}{\pi}}.
\label{eq18}\end{eqnarray}
}By replacing (\ref{eq18}) in (\ref{eq16}), in the asymptotic region, the variational energy
obtained for $\gamma<0$ is given by
{\small\begin{eqnarray}
E(\gamma,\beta)&\sim&
\frac{\beta}{8}-\frac{|\gamma|}{8\pi}  \sqrt{\frac{\beta}{\pi}}
\label{eq19}
.\end{eqnarray}
The minimization of the energy will give us $\beta_{min}= {|\gamma|^2/(4\pi^3)}$, with
the corresponding negative energy going asymptotically as $-\beta_{min}/8$.
By removing the $\beta-$dependence, for asymptotically large negative interactions, 
we obtain the energy and chemical potential as
{\small\begin{eqnarray}
E(\gamma)&\sim&
-\frac{{ |\gamma|^2}}{32\pi^3},\;\;\; \mu \sim -\frac{{3 |\gamma|^2}}{32\pi^3}
\label{eq20}
.\end{eqnarray}
We should also observe that the corresponding variational densities, given by $|f_v(\theta)|^2=
({\sin\theta})^\beta/[2\pi{\cal J}(\beta)]$, goes to a Dirac-delta function representation located
at $\pi/2$, when $\beta\to\infty$. This is being represented in panel (a) of Fig.~\ref{fig-04},
for a few values of the interaction parameter $\gamma\equiv\gamma_{11}+\gamma_{12}$,
in which variational results identified by the corresponding values of $\beta$ are being compared
with exact-numerical ones.
The results for the densities are quite representative of the deviations between exact and variational
results (deviations which are partially hidden in the observables as energy and
chemical potentials). Therefore, to enhance the deviation between variational and exact results,
the density peaks  are shown in panel (b) as a function of $\gamma$.
\begin{figure}[h]
    \centering
 \includegraphics[scale=0.3]{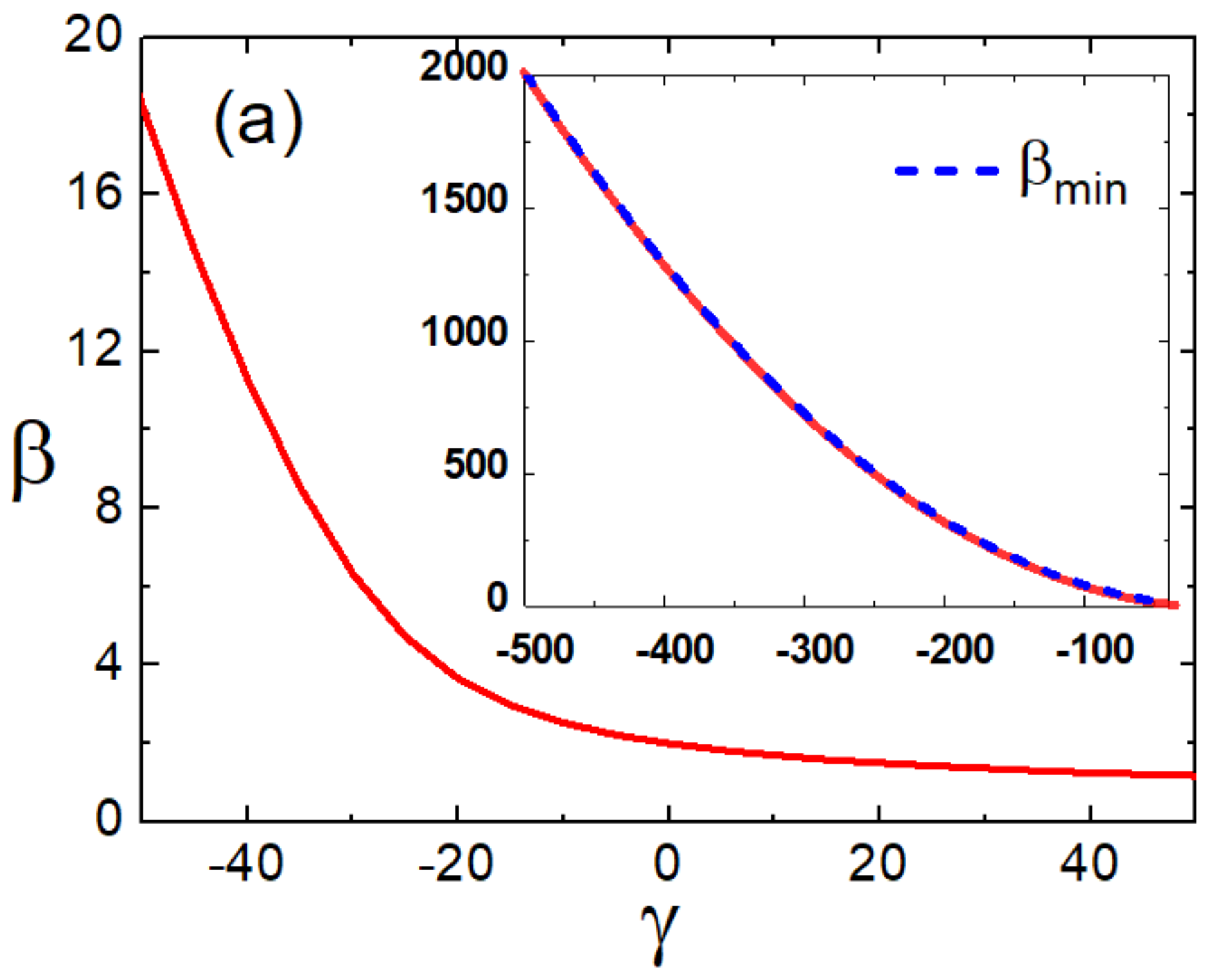}
   \includegraphics[scale=0.28]{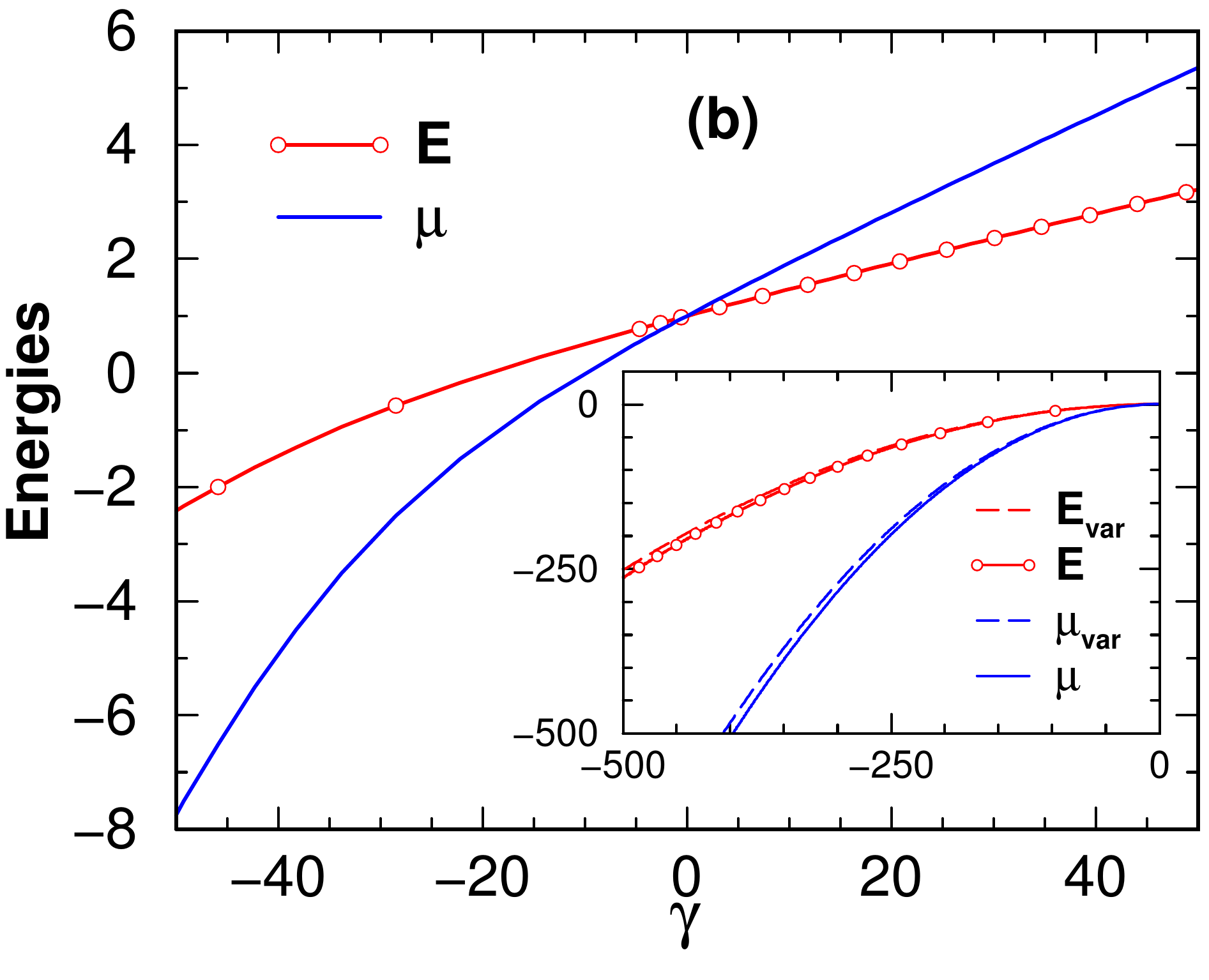}
   \vspace{-0.5cm}
    \caption{(Color on-line) 
    In terms of the summed interaction parameters $\gamma\equiv \gamma_{11}+\gamma_{12}$
    (with $\gamma_{11}=\gamma_{22}$), in panel (a) we have the minimization variational
    parameter $\beta$.
In panel (b), an almost perfect agreement between variational and exact-numerical results
are shown for the energies and chemical potentials. In both panels (a) and (b), the 
corresponding attractive asymptotic results are shown in the insets.
All quantities are dimensionless, with units defined in the text.
}
    \label{fig-03}
       \vspace{-0.2cm}
\end{figure}
\begin{figure}[h]
\hspace{0.5cm}
\includegraphics[scale=0.3]{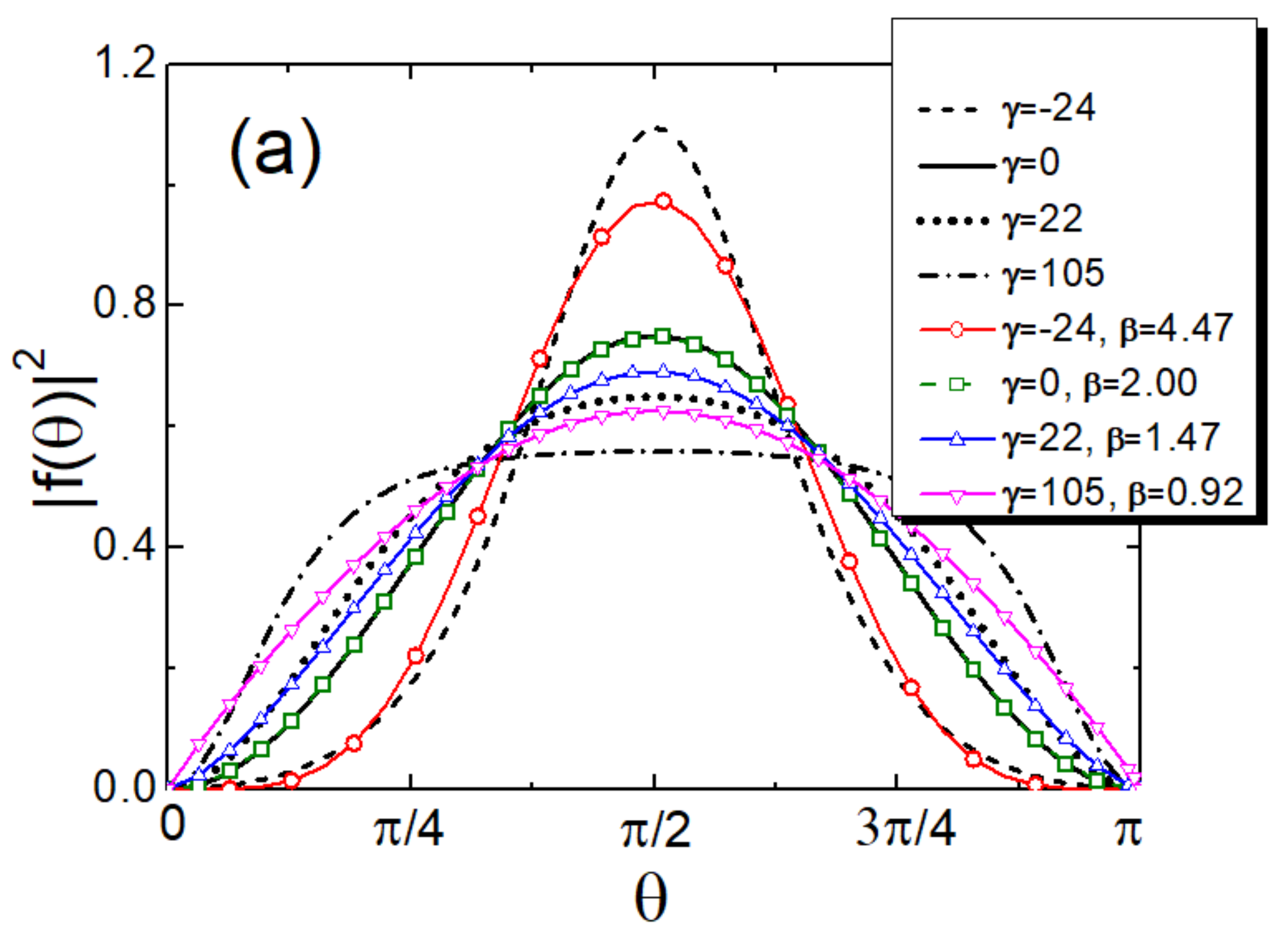}
\includegraphics[scale=0.3]{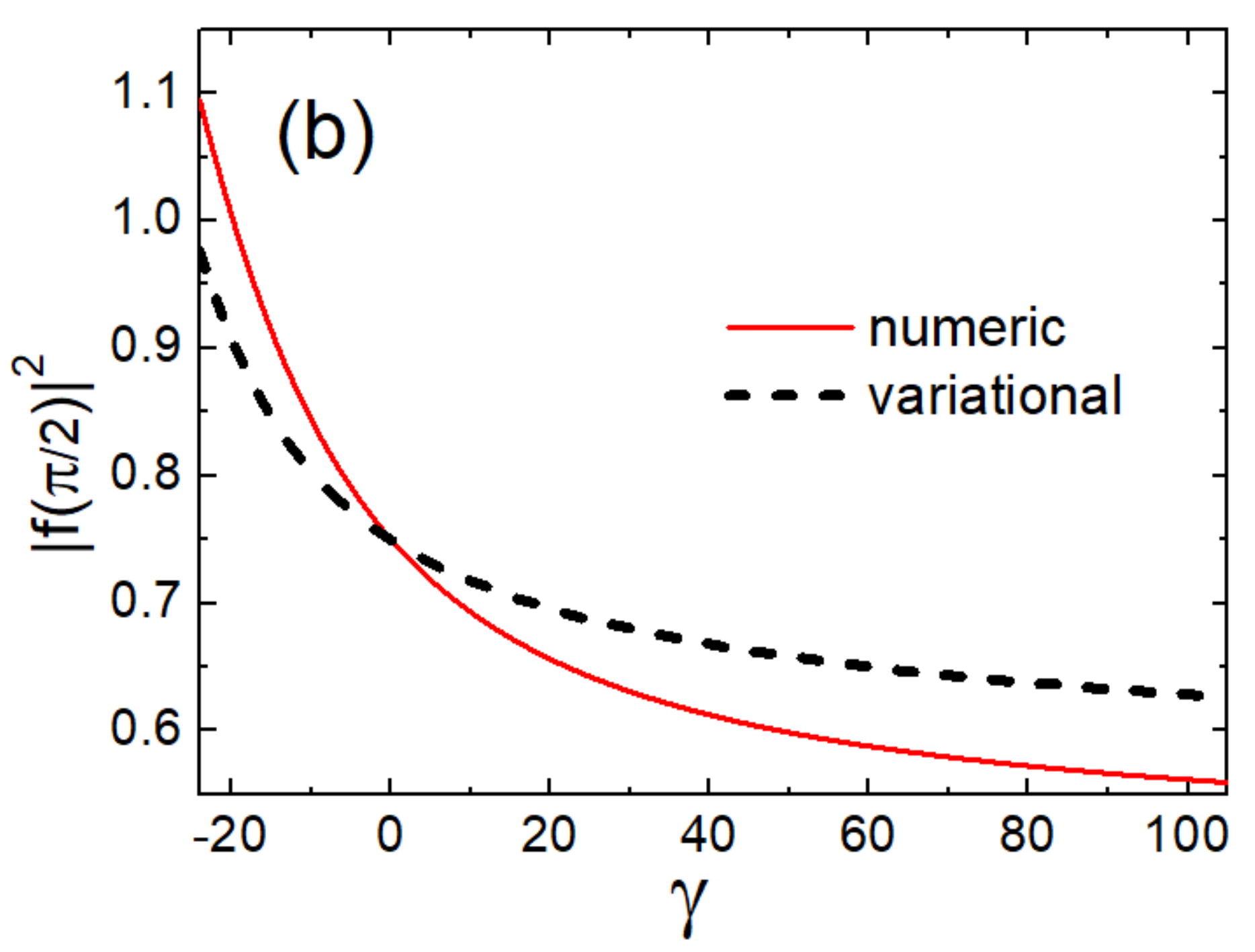}
    \caption{(Color on-line)  In the panel (a), considering four values of 
$\gamma$,    $|f(\theta)|^2$ variational results
(with $\beta$ indicated)  are being compared with the full numerical ones. 
In the panel (b), the density peaks [$1/{\cal J}(\beta)$, for the variational case] are shown 
as functions of $\gamma
\equiv \gamma_{11}+\gamma_{12}$ $(\gamma_{11}=\gamma_{22})$. 
All quantities are dimensionless, with units defined in the text.}
    \label{fig-04}
\end{figure}}
}

\subsection{Bogoliubov-de Gennes stability analysis}
Once verified the non-perturbed stationary solutions, their stability is probed by considering the
time-dependent analysis, considering small perturbations.
Due to the strict phase dependence of $\phi$ for the stationary state $\psi_{i0}$ in Eq.~\eqref{eq04},
it is convenient to follow  by expanding the perturbed states in an angular momentum basis, as 
done in Ref.~\cite{brtka2010}.
By separating the azimuthal dependence $\phi$ as a phase with quantum number $m$,
let us consider in Eq.~(\ref{eq06}), the infinitesimal time-dependent perturbations with the 
coefficients $u_{i\nu}$, $v_{i\nu}$ replaced by
{\small\begin{equation}
u_{im}(\theta,\phi)\equiv u_{im}(\theta) e^{{\rm i}m\phi}, \;\;\;
v_{im}^*(\theta,\phi)\equiv v_{im}^*(\theta)e^{{-\rm i}m\phi}.
\label{eq21}
\end{equation}
}By factoring the azimuthal dependence,  in the next, we follow by defining
$u_{im}\equiv u_{im}(\theta)$ and $v_{im}\equiv v_{im}(\theta)$.
The corresponding BdG equations,
which can be solved numerically for given integers $m \in \mathbb{Z}$, can be derived
by substituting the coefficients defined by Eq.(\ref{eq21}) in the corresponding differential nonlinear
equation~\cite{brtka2010}.

From that, by considering the chemical potentials $\mu_i$, with the corresponding
eigenfunctions $f_i(\theta)$,  as given in Eq.~\eqref{eq13}, the corresponding set of
coupled equations for ($u_1,v_1,u_2,v_2$), is given by the following BdG matrix:
{\footnotesize\begin{equation}
\begin{pmatrix}
{\cal D}_{1}^{+}      & \alpha_{11}   &   \alpha_{12}   &  \alpha_{12} \\
-\alpha_{11} &-{\cal D}_{1}^{-}   & -\alpha_{12}  &  -\alpha_{12} \\
\alpha_{21}  & \alpha_{21}&{\cal D}_{2}^{+}      &  \alpha_{22} \\
-\alpha_{21} & -\alpha_{21} &-\alpha_{22}   & -{\cal D}_{2}^{-}
\end{pmatrix}
\begin{pmatrix}
u_1\\ v_1\\ u_2\\ v_2\\
\end{pmatrix}=\omega \begin{pmatrix}
u_1\\ v_1\\ u_2\\ v_2\\
\end{pmatrix}.
\label{eq22}\end{equation}
}In the above, we are defining $\alpha_{ij}\equiv {\gamma_{ij}f_if_j}/{(2\pi)}$ and
${\cal D}_{i}^{\pm}\equiv \left(L^2_{s_i\pm m}\right)/2-\mu_i+
{(2\gamma_{ii}f_i^2+\gamma_{ij}f_j^2)}/({2\pi})\;\; (i\ne j).$
By given the values of $m$ and $s_i$, this system can be solved numerically.
Also, by considering Eq.~\eqref{eq02}, if one looks for a simplification of the matrix
elements, we can identify the following relations between the operators
${\cal D}_{i}^{\pm}$: One, which is in general given by
{${\cal D}_{1}^{+}-{\cal D}_{1}^{-} = {\cal D}_{2}^{-}-{\cal D}_{2}^{+}={2m}/{\sin^2\theta}$};
and another, valid for $\gamma_{11}=\gamma_{22}$, $\gamma_{12}=\gamma_{21}$:
${\cal D}_{1}^{+}-{\cal D}_{2}^{-} = {\cal D}_{1}^{-}-{\cal D}_{2}^{+}=\mu_2-\mu_1=0.$
With these relations, the four operators appearing in the diagonal part of (\ref{eq22}) can be 
reduced to just one operator, such that
{${\cal D}_{1}^{-}+\omega= {\cal D}_{1}^{+}+\omega-{2m}/{\sin^2\theta}$,
${\cal D}_{2}^{+}-\omega={\cal D}_{1}^{+}-\omega-{2m}/{\sin^2\theta}$, ${\cal D}_{2}^{-}+\omega=
{\cal D}_{1}^{+}+\omega$}.

Despite the general analytical treatment presented so far for the BdG equations, we restrict the
analysis to the stability of a mixture that was initially in a fully miscible configuration,
with $\gamma_{11} = \gamma_{22}$ and $f_1(\theta) = f_2(\theta) \equiv f(\theta)$.
Within the present assumptions, with $s_1=-s_2=1$, the numerical
approach requires first to solve Eq.~\eqref{eq13}, and then use the resulting
$f(\theta)$ in Eq.~\eqref{eq22}.
In Sec.~\ref{sec3} we have a particular case with $s_j = 0$ and $f_i(\theta)$ constant,
with the given numerical results for $\gamma_{11}=10$ shown in panel (b) of Fig.~\ref{fig-02}.

The numerical solution of Eq.~\eqref{eq13} is determined by using the Newton 
Conjugate-Gradient (NCG)
method, which is suitable for analytical continuation from a known solution
for a particular case~\cite{2009Yang}. The starting point is taken from the noninteracting case, 
which has the associated Legendre functions as general solutions, $P_{\ell}^{s}(\cos\theta)$, with
$s = \pm 1$ and $\ell = 1$, corresponding to the lowest energy level providing the HV
condition, which also implies that $\mu = 1$. 
Nevertheless, it is worth emphasizing that we have a full spectrum also when considering higher 
angular momentum $\ell$ that can be obtained using this analytic continuation procedure.

\section{Stability and dynamics of hidden vorticity states}\label{sec5}
In this section, we investigate the stability of the coupled stationary states (with their 
corresponding vorticity
established by $s_1=-s_2=1$) under small time-dependent oscillatory perturbation as 
given by \eqref{eq21}.
\begin{figure}[htb]
\centering
\includegraphics[scale=0.2]{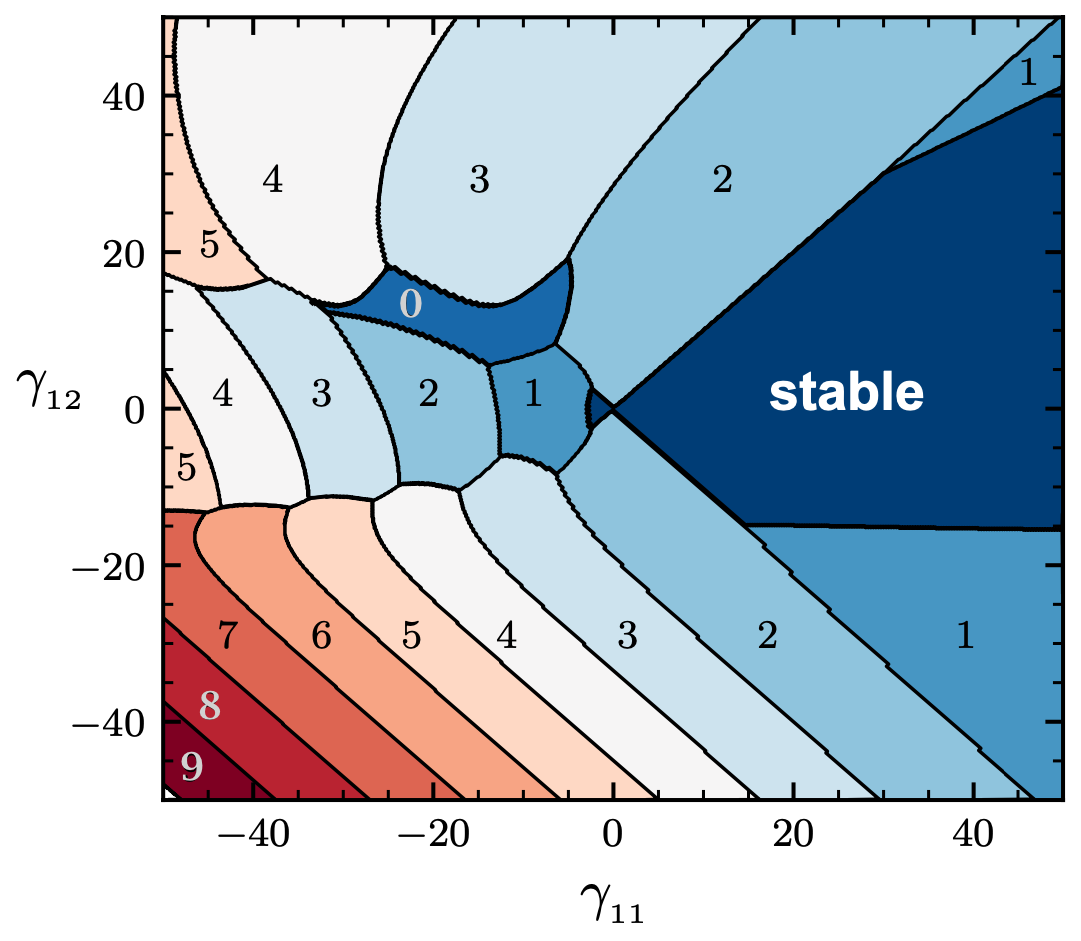}
\vspace{-.2cm}
\caption{(Color on-line) Stability diagram, in a phase space defined by the inter-species 
$\gamma_{12}$ versus the intra-species interactions $\gamma_{11}=\gamma_{22}$, for 
$|\gamma_{ij}|<50$.
The dark area is the stable region,  with the unstable ones dominated by perturbation modes with
$m$ up to 9 (as indicated). In all the cases, the vorticity is given by $s_1 = - s_2 = 1$. The 
$\gamma_{ij}$ interactions are dimensionless, with units defined in the text.
}
\label{fig-05}
\end{figure}
The instabilities are being verified for different modes of perturbations, which are numerically 
identified by the quantum number $m$ appearing in \eqref{eq21}. Therefore, systematically, by 
solving the corresponding GP formalism, we obtain the lowest order unstable modes.  
As considering the symmetry of the
solutions, which are identical for positive and negative values of $m$, in our following
analysis we are just referring to the positive values, with $m$ starting from 0.

\begin{figure}[!htb]
\includegraphics[width=7.5cm]{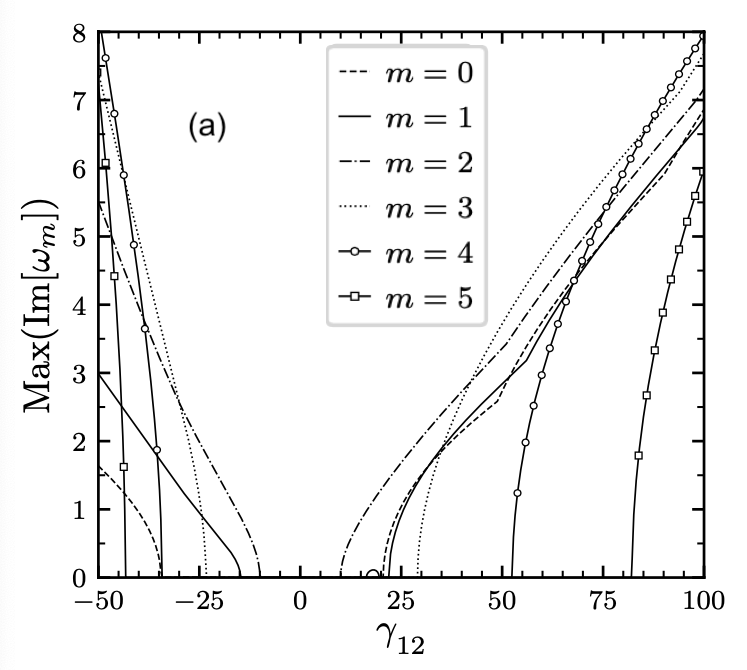}
\includegraphics[width=7.2cm]{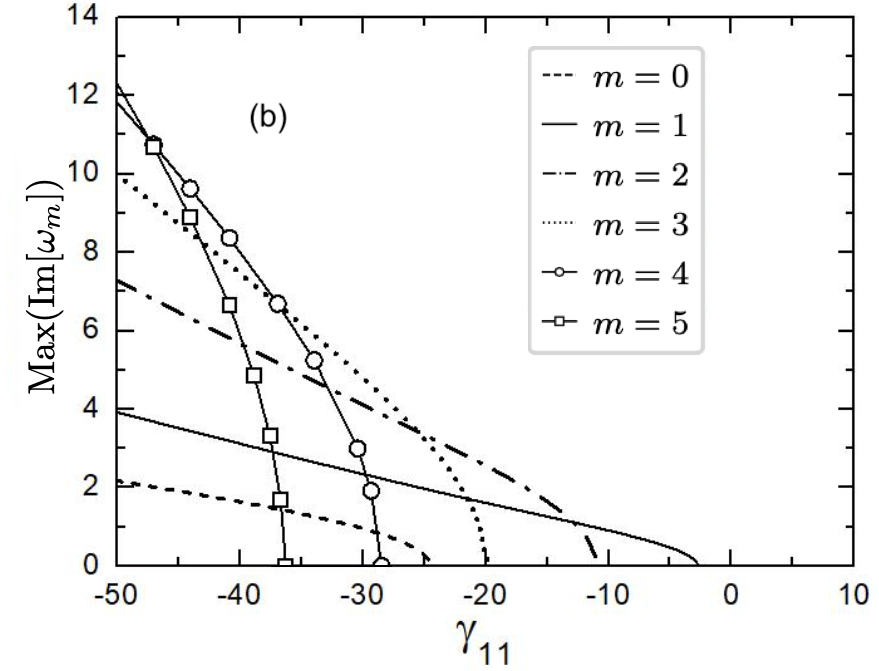}
\caption{
The above two panels, (a) and (b), refer to the imaginary spectrum of the BdG equations, given 
by the ${\rm Im}(\omega_m)$, with unstable modes up to $m=5$. They correspond to two lines 
of Fig.~\ref{fig-05}. The panel (a) is obtained by varying $\gamma_{12}$ with fixed 
$\gamma_{11}=10$;  with the panel (b), by varying $\gamma_{11}$ with fixed $\gamma_{12}=0$. 
As in Fig.~\ref{fig-05}, the vorticity is for $s_1 = - s_2 = 1$.
All quantities are dimensionless, with units defined in the text.
}
\label{fig-06}
\end{figure}
Concerning our general study for the stability of the system, we are summarizing the results in the
diagram shown in Fig.~\ref{fig-05}, in which the phase space is defined by the inter- and intra-species
interactions, with $\gamma_{11}=\gamma_{22}$. In this diagram, for the intra- and inter-species
interactions, we are assuming both possibilities that they can be repulsive $\gamma_{ij}>0$ or 
attractive $\gamma_{ij}<0$, varying from $-50$ up to $50$, with the only restriction that the 
intra-species interactions are identical for both species.
The diagram is indicating the stable and unstable regions with the corresponding predominant
modes, which is the one with the largest imaginary part, following the same procedure as we have
considered for the homogeneous case shown in Fig.~\ref{fig-01}.
However, by considering the inhomogeneous case, it is worth to emphasize the very different
behaviors depending on whether some of the interaction parameters can be negative.
In this case, the diagram is indicating the instability region, considering the dominant unstable
mode, from $m=0$ up to $m=9$. This highlights important features, since we can see how many modes
simultaneously can destabilize the system, as well as their magnitude, which is important by analyzing
the full numerical solution of the time-dependent problem.

In general, as verified in Fig.~\ref{fig-05}, many modes start to compete as $|\gamma_{12}|$ increases, 
even more rapidly for the attractive region. Moreover, it is also important to note from this diagram  
that the binary system can be stable mainly for $\gamma_{11}>0$, with repulsive and attractive 
inter-species $\gamma_{12}$ within some ranges.
Besides that, we can also observe a small stable interval for $\gamma_{11}<0$, when the inter-species
absolute value $|\gamma_{12}|$ is comparable with $\gamma_{11}$, which is related to the necessary 
energy at which we have the kinetic energy dominating, together with the interplay between attractive 
and repulsive non-linear interactions. As in the other regions, which are stable under small 
time-dependent perturbations, we have confirmed the stability of this particular region.  For example, 
by considering $\gamma_{11} = -2.0$ with $\gamma_{12} = 0$, the corresponding state remains stable 
for a larger time interval, going till $t = 100$, which we found enough for any manifestation of instability.

In Fig.~\ref{fig-06} our results are concentrated in two specific cases, in order to help elucidating
the results shown in the diagram, and expose the relevance of the different modes to generate the 
instabilities. In these two plots, we are considering the particular behavior of the maximum values of the 
unstable modes, for fixed values of one of the interactions.
In the panel (a) we fix the intra-species interactions, with $\gamma_{11} = 10$,  with the inter-species 
interaction $\gamma_{12}$ varying in a larger interval than the one shown in Fig.~\ref{fig-05}, from $-50$ 
up to $+100$. The stable regions are clearly identified as the ones for $|\gamma_{12}|\le\gamma_{11}=10$, 
in this case, with the first dominant unstable mode being for $m=2$. The competing behaviors of all the 
unstable modes, up to $m=5$, are shown in this panel, at which the dominant modes (for the instability) are 
the ones with the largest values for the $Im(\omega_m)$. In panel (b) of Fig.~\ref{fig-06}, we present the 
corresponding spectrum for the case $\gamma_{12}=0$, in which the system is uncoupled, such that both 
species $1$ and $2$ have the same spectrum, considering that we are assuming $\gamma_{11}=\gamma_{22}$. 
This figure, more than indicating the stable regions shown in Fig.~\ref{fig-05}, also clarifies how the different 
modes contribute to the instability.

From the initial form of the perturbations, given by Eqs.~\eqref{eq06} and \eqref{eq07}, we interpret
imaginary values of $\omega$ as an exponentially growing perturbation, which in turn shows that, after some
evolution period, the perturbations should not be assumed small in comparison with the condensate
wave functions (\ref{eq04}), as initially assumed to obtain the BdG equations. This implies that any
initial perturbation different from zero will drastically change the condensate state after a sufficiently 
long time. To analyze such effect, we just use the respective stationary state obtained numerically as the initial
condition in the full time-dependent problem represented of Eq.~\eqref{eq01},
whereas any exponentially raising perturbation is triggered by the numerical noise.

\begin{figure}[t]
    \centering
    \includegraphics[scale=0.8]{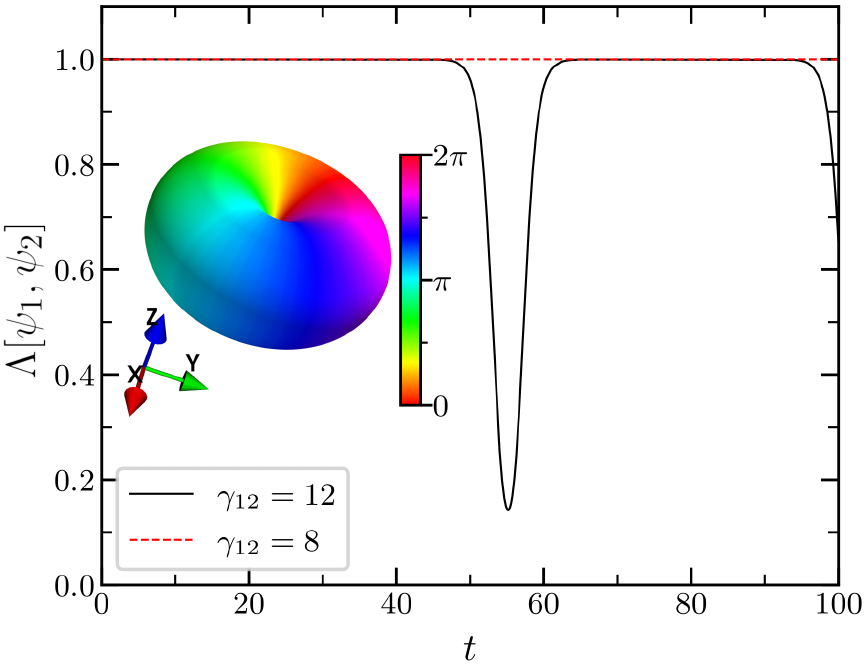}
\caption{(Color on-line) Time evolution of the two-species overlap for a hidden vorticity mixture state with 
$s_1 = -s_2 = 1$ and $\gamma_{ii} = 10$, in a
stable ($\gamma_{12} = 8$, with dashed line) and an unstable ($\gamma_{12}=12$, with solid line) region.
At $t \approx 48$, with $\gamma_{12}=12$, the miscibility suffers a short pulsed change from the initial
condition, which is verified to be periodic in a longer-time interval (See Fig.~\ref{fig-08}).
The radial density 3D representation is for the initial condition of $|f_1(\theta)|^2$ with $\gamma_{12} = 12$,
in which the phases around the surface are mapped to colors (The corresponding 1D plot of $|f_1(\theta)|^2$
is in Fig.~\ref{fig-04}, with $\gamma_{11}+\gamma_{12}=22$).
All quantities are dimensionless, with units defined in the text.
}
\label{fig-07}
\end{figure}
As a main measure to track both species density behavior, we introduce a functional for the
miscibility of both time-dependent densities
$|\psi_i|^2\equiv|\psi_i(\theta,\phi,t)|^2$, defined by
\begin{equation}
    \Lambda[\psi_1, \psi_2] \equiv
\frac{\left[ \int \mathrm{d} \Omega |\psi_1|^2 |\psi_2|^2 \right]^2}
{\int \mathrm{d}\Omega |\psi_1|^4 \int \mathrm{d}\Omega |\psi_2|^4} ,
\label{eq23}
\end{equation}
which is one for complete overlap of the densities (miscible mixture); reducing to zero
when the coupled system is completely immiscible. In the following we present our
main results, exemplified by the case with intra-species interactions fixed at
$\gamma_{11}=\gamma_{22}=10$. For the inter-species interaction, we present results
for repulsive and attractive cases. The general diagrams presented in Fig.~\ref{fig-05} indicate
that similar features could be verified for other values of the interactions.

In the next section, our main results are illustrated with the analysis of the
dynamics of few representative cases considering repulsive and attractive
inter-species cases. We are
mainly focused on the cases that we have repulsive intra-species
interactions.

\subsection{Dynamics of unstable states - repulsive inter-species case}
\begin{figure}[t]
    \centering
        \includegraphics[scale=0.8]{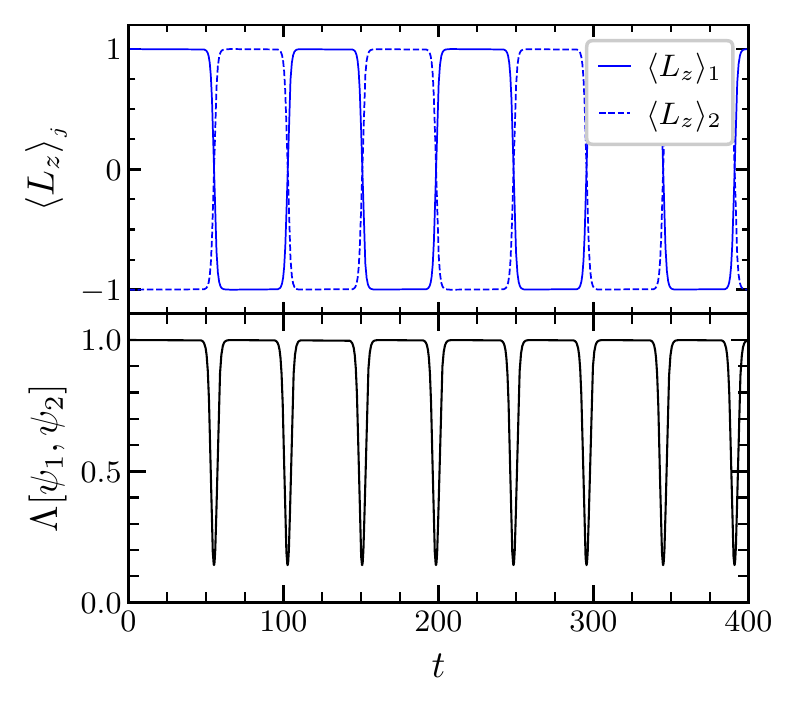}
    \caption{For the same conditions given in Fig.~\ref{fig-07}, we show the corresponding 
    long-time evolution (up to $t = 400$) of the two-component angular momenta, 
    $\langle L_z\rangle_i$ (upper panel), with the associated density overlap $\Lambda$ (lower panel). 
    A clear periodic behavior is verified for both.
    All quantities are dimensionless, with units defined in the text.
    }.
    \label{fig-08}
\end{figure}

The result for time evolution of the miscibility functional (\ref{eq23}) is presented in
Fig.~\ref{fig-07}, considering a time interval $t<100$, in which
we are comparing the time evolution of two states subject to different stability conditions.
As verified, when using $\gamma_{12} = 8$ the overlap between the two states remains complete.
However, it appears an unstable branch for $\gamma_{12} > 10$.
As can be noted, in agreement with the BdG prediction, for $\gamma_{12} = 12$ case the overlap
changes drastically near $t \approx 48$, indicating that indeed some perturbation became relevant to
the condensate wave function, growing from initial numerical finite precision. Meanwhile, there is no
change in the density profile for both species for $\gamma_{12} = 8$ as expected since it is stable
against small perturbations.
Near the final time observed, $t \approx 90$ the overlap starts to change again after a period remaining
in the initial value. A sketch of the initial state is also provided as a surface plot where the radius of the
surface was taken as $|f_1(\theta)|^2$, that is equal to $|f_2(\theta)|^2$, but their phases have different
orientations as $s_1 = -s_2 = 1$, which are displayed in colors over the surface.
For a more complete picture, we have verified the time evolution of the miscibility for a longer time 
interval, shown in Fig.~\ref{fig-08}.

Analyzing Fig.~\ref{fig-07} carefully, for this initial condition we see that after a brief period, the overlap 
functional $\Lambda$ becomes smaller than 1, approximately for $t \in [45, 65]$, it stands for a long 
period in its initial value, up to $t \approx 92$, when it starts to decrease again.  Therefore, naturally 
arises the question of whether this behavior is periodic or not. In Fig.~\ref{fig-08}, we can confirm the 
periodic behavior not only for $\Lambda$, but also for the angular momentum of both species in a 
long time dynamics. Nevertheless, the period of the angular momentum of each species is twice the 
period of $\Lambda$ and it reveals an interesting feature as the species exchange their momenta 
between $\pm 1$ as $\Lambda$ returns to 1.

The oscillating period inferred from Fig.~\ref{fig-08} was $\tau = 47\pm1$ for
$\Lambda$. To obtain the period, we used the $\Lambda < 0.2$ points as suggested by the minimum
in Fig.~\ref{fig-07}, and computed time instants where the derivative vanished, from which
we computed the average and standard deviation (explaining the $\pm 1$ in $\tau$). 
However, it is worth emphasizing that this value depends on how the instability is triggered. 
In our case,   it is due to the finite precision of numerical calculations.

\begin{figure}[t]
    \centering
    \includegraphics[scale=0.8]{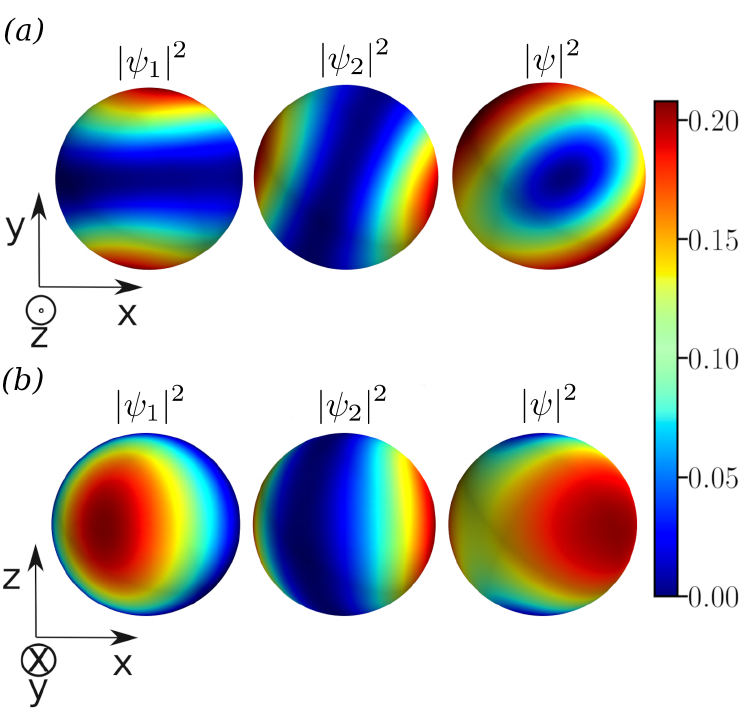}
    \caption{(Color on-line) Density plots for both species and average density at $t = 55$ of the dynamics 
    presented for $\gamma_{12} = 12$. Frame (a) displays the images with $z$ axis pointing
    outwards
    the page while frame (b) provide a $90^o$ rotation with respect to (a), with $y$ axis pointing
    inwards the page.
    All quantities are dimensionless, with units defined in the text.}
    \label{fig-09}
\end{figure}
In Fig.~\ref{fig-09} we provide some snapshots of the density at $t = 55$ of both species
illustrated by colors in a spherical shell. As can be seen, not only the BdG prediction can be
confirmed as the instability mode grew and changed completely the initial density profiles, that
were independent of $\phi$ angle,
but also the density for both species breaks up in $2$ disconnected pieces along the $\phi$ angle in
the spherical shell, which corroborates with a superposition of modes with spatial frequency $m = \pm 2$.
Specifically in this case, corresponding to $\gamma_{12} = 12$ in Fig.~\ref{fig-06}(a), only 
$m = 2$ contributes, explaining why we should expect only two pieces in the dynamical breakup,
although it is not a rule when more than one mode is unstable.

\begin{figure}[t]
    \centering
    \includegraphics[scale=0.54]{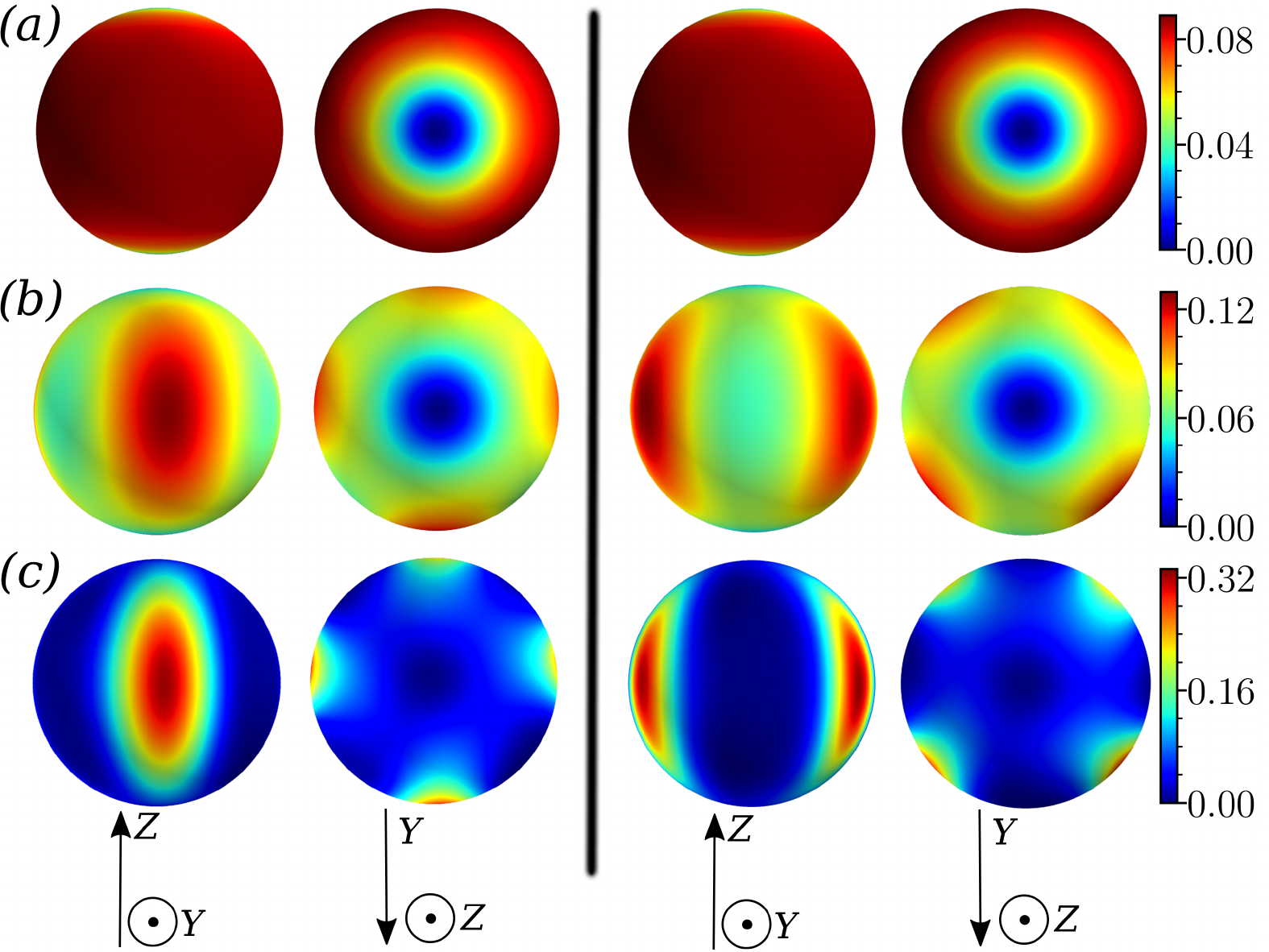}
    \caption{(Color on-line) Density time-evolution for the coupled system with $m=4$
    dominant unstable mode. The first two columns are for the species 1, with the remaining
    two columns for the species 2, considering two visualization angles, as indicated in the
    graphical legends at the bottom. The upper row (a) represents the two densities at the
    initial stage $t=0$ (when both species are completely mixed, $|\psi_1|^2 = |\psi_2|^2$).
    The middle row (b) is for an intermediate time $t = 3.85$, with the system partially immiscible with four
    maxima being distinguished already in the sphere. In the lower row (c), $t = 4.15$, we have already a final
    complete immiscible configuration with the density peaks well localized on the sphere, at different positions.
    The interaction parameters are $\gamma_{11}= \gamma_{22} = 10$ and $\gamma_{12}
    = 95$, with the initial condition obtained by the lowest  energy state with $s_1 = - s_2 = 1$.
    The time instants are dimensionless, with units given in the text.}
    \label{fig-10}
\end{figure}

Finally, we explore another case with large inter-species interaction in real-time evolution, using
$\gamma_{12} = 95$ still for $\gamma_{11} = \gamma_{22} = 10$ in Fig.~\ref{fig-10}. The two
first columns refer to the first species densities, while the last two correspond to
the second, in different view angles as denoted by the axes legend at the bottom.  At the initial instant,
both species share the same density profile, which can also be seen as a 1D plot in
Fig.~\ref{fig-04}, vanishing at the top due to HV condition. In a second instant $t = 3.85$ in the row (b)
we already can see the formation of 4 localized peaks around the sphere along $\phi$ direction,
though the contrast is not so prominent. In a third instant $t = 4.15$ let clear the immiscibility
of the mixture with 4 localized and narrow peaks. The color scale showed for each instant
provide a quantitative comparison for the density clustering of the 4 pieces.

\begin{figure}[h]
    \centering
    \includegraphics[scale=0.9]{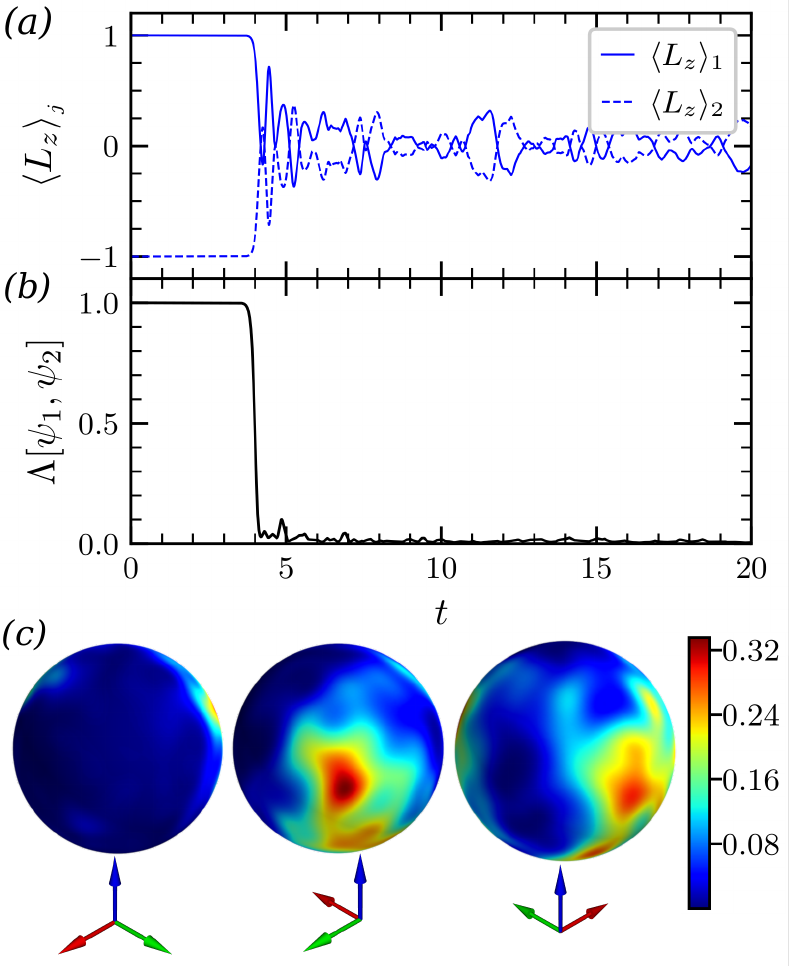}
    \caption{(Color on-line) The time evolution for $0<t<20$ is shown for $\langle L_z\rangle_i$ (a) and for the
    overlap $\Lambda$ of the two components (b), considering
    the unstable hidden vorticity state $s_1 = -s_2 = 1$ with $\gamma_{ii} = 10$ 
    and $\gamma_{12} = 95$. A drastic variation if verified  
    close to $t = 4$. In (c), three different angle views of the density distribution are selected 
    for the species 1 at $t=16$.
    All quantities are dimensionless, with units defined in the text.
    }
    \label{fig-11}
\end{figure}

There is a sharp contrast between the two cases evaluated in Fig.~\ref{fig-09} and Fig.~\ref{fig-10}
in the time elapsed until the unstable mode becomes dominant,
in the first case at $t \approx 48$ and in the second $t \approx 3.8$. This is explained by the
magnitudes of the imaginary part of the BdG eigenvalues provided in Fig.~\ref{fig-06}, and more
rapidly the mode will destabilize as larger is
$\gamma_{12}$, though we cannot expect any proportional relation as the initial numerical
inaccuracy that triggers these modes is hard to estimate. However, Fig.~\ref{fig-06} points out
that the region $\gamma_{12} = 95$ has many competing modes while in
$\gamma_{12} = 12$ there is only the $m = 2$ mode as unstable, and in contrast to
the periodic behavior observed in the latter case, in Fig.~\ref{fig-11} we can see that no
clear pattern can be detected. We attribute the unrecognizable pattern in the time evolution
as a consequence of many modes excitation.

\subsection{Dynamics of unstable states - attractive inter-species case}

The case of attractive inter-species interaction was also explored in our study, as shown in the diagram given
in Fig.~\ref{fig-05} and by the stability spectrum shown in Fig.~\ref{fig-06}. In contrast to the repulsive cases,
when the two species are prone to breaking down into immiscible pieces, the attractive interaction should
maintain the overlapping densities as initially prepared in the stationary state, as being energetically 
favorable.  Therefore, they shall maintain the miscibility, with the unstable modes indicating that it
will not be uniform in the azimuthal $\phi$ direction.

\begin{figure*}[t]
    \centering
    \includegraphics[scale=0.8]{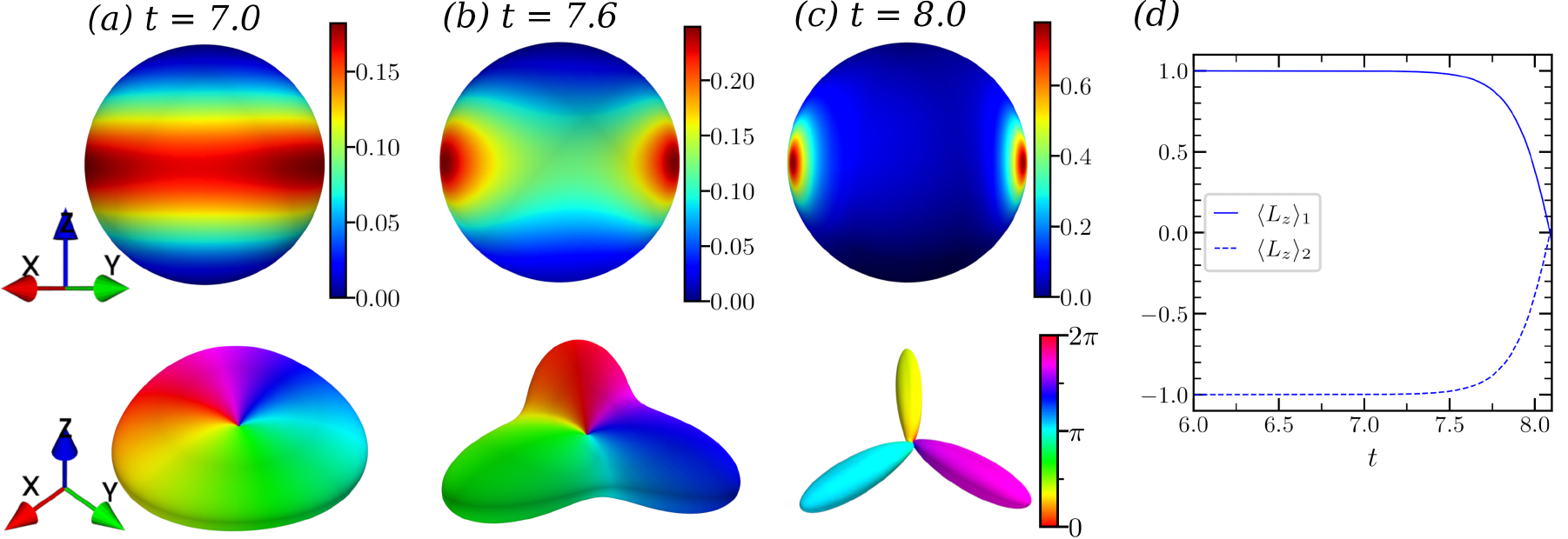}
    \caption{(Color on-line) Density plots $|\psi_1|^2 = |\psi_2|^2$, in the 
    inter-species attractive case with $\gamma_{12}=-34$ and $\gamma_{ii}= 10$,
    within a completely miscible configuration ($\Lambda=1$).
    Three-time instants close to the collapse are shown, for $t=$7.0 (a), 7.6 (b), and 8.0 (c), 
    with the instability dominated by $m=3$. 
    In the upper row, they are mapped to colors on the sphere surface (central visualization at 
    $\theta=\pm\pi/2$). The maxima are in three $\phi$ positions, 
     which are best verified in the lower row with $z$-axis tilted by $45^{\circ}$,
    where the densities have radial representation with the phases mapped to colors. 
    In panel (d), we have the two-species time evolution $\langle L_z\rangle_i$, close to the collapse.  
    The initial condition is for the lowest energy state with $s_1 = - s_2 = 1$.  
    All quantities are dimensionless, with units given in the text.
In the Supplemental Material~\cite{video}, a movie illustrates the corresponding 
full-time evolution of the density till the collapse.
}
    \label{fig-12}
\end{figure*}

In Fig.~\ref{fig-12} we provide an example of real-time evolution of unstable HV
state with $\gamma_{12} = -34$ and $\gamma_{11} = \gamma_{22} = 10$.
During all the time observed, both species have complete overlap, with $\Lambda = 1$.
The instability starts growing drastically for $t \gtrsim 7$, being dominated by the $m = 3$ mode,
which can also be followed by looking at our diagrammatic representation given in Fig.~\ref{fig-05},
corresponding to the largest imaginary part in the BdG spectrum. More closely,
the dominance of the mode $m=3$ for  $\gamma_{12} = -34$ is also shown in Fig.~\ref{fig-06}.
The dominance of this mode implies in the densities being broken into three pieces around the sphere.
From the time instant $t = 7$, when the densities start to be accumulated at different positions
on the sphere, we show snapshots of them at $t = 7.6$ and $t = 8.0$.
In the upper row, with $|\psi_1|^2 = |\psi_2|^2$, both densities are shown by color-density plots.
In the lower row, the full wave function is represented by using a radial surface plot for the densities,
with a color mapping for the phase, with the $z$-axis tilted.
The numerical solution could be evaluated up to $t = 8.1$, as at this instant the condensate wave functions
start to become singular in the numerical grid, breaking the energy and norm conservation.
Despite both methods of visualization are complementary, especially at $t = 8.0$ when the condensates
are about to collapse, the radial plot is more suitable to see the three peaks. The two-species 
angular momentum values,  $\langle L_z\rangle_i$, are also provided for completeness, with both 
being zero at the collapse instant.

In this case shown in Fig.~\ref{fig-12}, the attractive inter-species interaction $\gamma_{12} = -34$ is 
dominating against the repulsive intra-species one, given by $\gamma_{11} = 10$. Therefore, effectively, we 
have an overall attractive interaction with the system collapsing. Indeed,  in the beginning we observe 
the instability of the mode $m=3$ which split the condensate into three pieces.
Next, after a short time interval,  the effective attraction shrinks the localized densities till the collapse.

\section{Conclusions and perspectives}\label{sec6}

Summarizing our main outcome,  we provide a dynamical stability study of a binary Bose-Einstein 
condensed mixture trapped on the surface of a rigid spherical shell, exploring the miscibility
of the system with and without vortex charges.
For that, the initial stationary solutions are treated by using the usual GP mean-field approach, 
within two possible configurations. First, with both species within a homogeneous non-vorticity 
mixture; next,  when the initial configuration is in the lowest non-interacting stationary states,  
with opposite charge vorticity.
The stationary solution study is supplemented by a variational analysis
for the specific case in which the non-linear system is with both species at the lowest level
($\ell=1$) that allows the existence of hidden vorticity.

In our approach to explore the stability of the system, we consider the Bogoliubov-de Gennes 
method, in which the stationary non-linear formalism is submitted to a linearization, with  
small time-dependent perturbation modes. The analysis of both the cases, 
with and without vortex charges, is followed by computing the critical points 
at which the system becomes unstable.

The relevance of the different modes to generate the instabilities is being exposed in the 
sample results presented in the panels of Fig.~\ref{fig-06}, which are obtained numerically for 
a large range of repulsive and attractive interactions. From this kind of analysis, a complete diagram is
provided in Fig.~\ref{fig-05}, in the phase space defined by the inter- and intra-species interactions,
where we can verify the stable regions, together with the dominating unstable modes. In all these
cases, we assume opposite charge vorticity in the lowest initial stationary states.

The study of the dynamics of the mixture is been represented by the 
time evolution of the miscibility, given by the overlap of the densities, 
together with the angular momentum distribution of the two species. 
For this purpose, we select a few cases to characterize stable and different 
unstable regimes. In the selected examples, where we kept fixed and repulsive 
the intra-species interactions, we illustrate the time-evolution of the coupled 
densities distributed in the bubble surface,  by considering immiscible and 
miscible regimes, which are, respectively, given by repulsive and
attractive inter-species interactions.

In the given illustrations, where we fix the intra-species interaction to be repulsive, 
we observe that 
the dominant instability mode determines the number of parts that the homogeneous 
state breaks down.
As shown, for sufficient high interaction, although the total angular momentum remains 
conserved, the individual vorticity can be lost.
In the case of attractive inter-species interactions, we first observe the existence of 
a stable region, which occurs due to the geometry.  Also verified in this attractive 
case, is the split of the system according to the most unstable mode, before  
the occurrence of the collapse.

Finally, we understand that the present study can be relevant in order to establish
initial parameters for experimental tests and realizations, as well as for more involved
theoretical approaches in which the miscibility of different kind of particles can be 
modified by their specific characteristics. 
Apart of possible experimental setup difficulties for realizations in microgravity conditions, 
a perspective theoretical investigation should be to extend the present study to 
the case of strongly mass-imbalanced mixtures, such as with $^{87}$Rb and 
$^{133}$Cs~\cite{2011McCarron,2011Lercher}, which have been explored in cold-atom
experiments. In view of previous studies~\cite{2020Kumar},  binary systems with
strong mass differences are expected to impact on the results we have presented, 
being of interest to verify how the stability and density distributions are affected.
Therefore, by considering coupled systems with identical
masses, the immediate possible applications of our analysis could be when considering 
binary mixtures of the same isotope with different internal spin states,
as being considered in Refs.~\cite{1996Ho,1997Myatt}, or in the 
case of a mixture with two close atomic isotopes, as $^{85}$Rb and  $^{87}$Rb, which 
should present similar results as the ones observed for identical mass mixtures. 
Another unavoidable future investigation refers to energetic instabilities, expected to 
emerge in the condensates with quantized vortices in spherical geometry, which can be verified 
by considering a time-dependent dissipation mechanism related to condensate 
interactions with the thermal cloud.

\appendix
\section{Numerical method for time evolution}\label{appendix}
The numerical approach to solve Eq.~\eqref{eq01} involves a combination of techniques.
First, a split-step method is used to separate the evolution of nonlinear and linear parts. 
Thus, from \eqref{eq01}, at an arbitrary instant, we take a function 
$\zeta_i\equiv\zeta_i(\theta,\phi,t)= \psi_i$, which satisfy
\begin{equation}
    \frac{\partial}{\partial t} \zeta_i = \left( \sum_{k} \gamma_{ik} |\psi_k|^2 \right) \zeta_i .
    \label{eqA1}
\end{equation}
The nonlinear part is propagated in small time steps with direct exponentiation,
since both species densities are considered frozen and treated as static potentials.

The linear part demands some care to
 handle the boundary conditions at the poles. A suitable approach is to work on the Fourier
 transformed space in the $\phi$-direction, which simplifies the Laplacian~\eqref{eq02}
 for each mode in the series. Thus from $\zeta_i$, we introduce
\begin{equation}
    \zeta_i = \sum_k e^{{\rm i} k \phi} \tilde{\zeta}_{ik} ,
\label{eqA2}\end{equation}
for the given time instant, whereas the Fourier weights $\tilde{\zeta}_{ik}$ are functions of
$\theta$ and time. Thereafter, from the linear part of Eq.~\eqref{eq01} we have the following
time-dependent equation for each mode $k$
\begin{equation}
     \frac{\partial}{\partial t} \tilde{\zeta}_{ik} = -\frac{1}{2}
\left[\frac{1}{\sin\theta} \frac{\partial}{\partial \theta}
\left(\sin\theta\frac{\partial}{\partial \theta}\right)
-\frac{k^2}{\sin^2\theta}
\right]
     \tilde{\zeta}_{ik} .
    \label{eqA3}
\end{equation}
Within this approach, we can now clearly
introduce the boundary conditions at the poles, which depend on the frequency mode $k$
along the $\phi$ direction as
\begin{eqnarray}
&&\left.\tilde{\zeta}_{ik}\right|_{\theta=0, \pi}= \delta_{0,k} ,\;\;\;
\left.\frac{\partial  \tilde{\zeta}_{i0}}{\partial\theta}\right|_{\theta=0, \pi} = 0\label{eqA4}.
\end{eqnarray}

Once the boundaries are set appropriately, the finite differences Crank-Nicolson semi-implicit
method was used to integrate the equations for each mode $k$ in Eq.~\eqref{eqA3}. After
the solutions are determined for the next step for the coefficients, they are transformed
back to the spatial $\phi$ coordinate.

In summary, the split-step method requires a time step and a discretization grid, then
Eq.~\eqref{eqA1} is propagated
half time step, then the linear part is solved in Eq.~\eqref{eqA3} for an entire time step
going forward and backward of the Fourier space, and finally, the nonlinear part is
propagated by another half time step, with the initial condition taken from the
resulting linear part propagation.

\section*{Acknowledgements}
The authors thank discussion with Profs. E.J.V. Passos and A. F. R. T. Piza. We also acknowledge the
Brazilian agencies Funda\c{c}\~ao de Amparo \`a Pesquisa do Estado de S\~ao Paulo (FAPESP)
[Contracts 2018/02737-4 (AA), 2017/05660-0 (LT), 2016/17612-7 (AG)], Conselho Nacional
de Desenvolvimento Cient\'\i fico e Tecnol\'ogico [Procs. 304469-2019-0(LT) and
306920/2018-2 (AG)] and Coordena\c c\~ao de Aperfei\c coamento de Pessoal de N\'\i vel
Superior [Proc. 88887.374855/2019-00 (LB)].

\end{document}